\pdfoutput=1
\newcommand*{\ATLASLATEXPATH}{latex/}
\documentclass[cernpreprint,texlive=2016,txfonts,UKenglish]{latex/atlasdoc}
\usepackage{etex}
\reserveinserts{10}
\usepackage{\ATLASLATEXPATH atlaspackage}
\usepackage[maxfloats=50]{morefloats}
\usepackage{\ATLASLATEXPATH atlasbiblatex}

\usepackage{\ATLASLATEXPATH atlasphysics}

\usepackage{floatrow}
\graphicspath{{logos/}{figures/}}

\usepackage{Ztaupol-defs}
\mathchardef\mhyphen="2D

\AtlasTitle{Measurement of $\tau$ polarisation in $Z/\gamma^{*}\rightarrow\tau\tau$ decays in proton--proton collisions at $\sqrt{s}=8$ TeV with the ATLAS detector}

\AtlasAbstract{%
This paper presents a measurement of the polarisation of $\tau$ leptons
produced in $Z/\gamma^{*}\rightarrow\tau\tau$ decays which is performed
with a dataset of proton--proton collisions at $\sqrt{s}=8$~TeV, corresponding
to an integrated luminosity of 20.2 fb$^{-1}$ recorded with the ATLAS detector at the LHC
in 2012. The $Z/\gamma^{*}\rightarrow\tau\tau$ decays are reconstructed from a
hadronically decaying $\tau$ lepton
with a single charged particle in the final state, accompanied by a $\tau$
lepton that decays leptonically.
The $\tau$ polarisation is inferred from the relative fraction of energy
carried by charged and neutral hadrons in the hadronic $\tau$ decays.
The polarisation is measured in a fiducial
region that corresponds to the kinematic region accessible to this analysis.
The $\tau$ polarisation extracted over the full phase space within the $Z/\gamma^{*}$ mass range of
66 $ < m_{Z/\gamma^{*}} < $~116~GeV is found to be $P_{\tau} =-0.14
\pm 0.02 (\text{stat}) \pm 0.04 (\text{syst})$.  It is in agreement with
the Standard Model prediction of $P_{\tau} =-0.1517 \pm 0.0019$, which is obtained
from the ALPGEN event generator interfaced with the PYTHIA 6 parton shower modelling and the
TAUOLA $\tau$ decay library.
}

\author{The ATLAS Collaboration}

\AtlasRefCode{STDM-2015-18}

\PreprintIdNumber{CERN-EP-2017-172}

\hypersetup{pdftitle={ATLAS document},pdfauthor={The ATLAS Collaboration}}

\begin{document}

\maketitle
\section{Introduction}
\label{sec:intro}

The $\tau$ lepton plays an important role in the physics programme of the Large Hadron
Collider (LHC). It is used to identify and measure electroweak and top quark
production processes as well as in searches for new physics beyond the Standard
Model. Since the $\tau$ leptons decay before exiting the ATLAS detector volume, their
polarisation can be measured.
 
The $\tau$ polarisation, $P_{\tau}$, is the asymmetry of the cross-section for
positive ($\sigma_+$) or negative ($\sigma_-$) helicity $\tau$ lepton
production, defined by:
\begin{equation}
\Ptau = \frac{\sigma_+ - \sigma_-}{\sigma_+ + \sigma_-}
\label{eqn:ptau}
\end{equation}
for the $\tau^{-}$ lepton. It is a measure of the degree of parity
violation in the interaction producing the $\tau$ leptons and therefore it provides insight into
the nature of its Lorentz structure. The positive
(negative) helicity states and right-handed (left-handed) chiral states
coincide in the relativistic limit assumed here.\footnote{The $\tau$ helicity states are
experimentally accessible, in contrast to chiral states, as the kinematic distributions of the $\tau$ decay products are sensitive to the spin
of the $\tau$ lepton. The inaccuracies resulting
from the assumption of helicity--chirality equivalence are negligible for the $\tau$ decays studied in this
analysis. } Due to nearly exact $\mathit{CP}$ invariance in $\tau$ decays, the kinematic
distributions for left-handed (right-handed) $\tau^{+}$ follow those of
right-handed (left-handed) $\tau^{-}$. Therefore, in this paper only one of 
the equivalent $\mathit{CP}$ states is mentioned at a time with the other being implicitly assumed.
Any possible differences are negligible for the measurement described in this paper. 

The $\tau$ polarisation in $Z\rightarrow\tau\tau$ decays was first measured at
LEP in electron--positron annihilation events at the $Z$ boson pole. 
The experiments at LEP published the \Ptau spectrum as a function
of the angle between the directions of the $\tau^{-}$ lepton and the $e^{-}$ beam~\cite{Ptau_LEP_new}.
The most precise value of the average $\tau$ polarisation was
obtained in the combination of LEP results and presented in terms of the $\tau$
production asymmetry, $A_{\tau}$, which, by convention, has reversed sign with
respect to the polarisation and contains small ($\mathcal{O}(0.005)$) corrections for the interference
between the $Z$ boson and photon propagators as well as for the pure photon contribution.
The asymmetry value obtained in the
combination is $A_{\tau} = 0.1439 \pm 0.0043$~\cite{Ptau_LEP_new}.

The measurement presented in this paper provides a complementary constraint on the $\tau$ polarisation in
decays of $Z/\gamma^*$ that are produced via a $qqZ$ vertex in proton--proton collisions as the quark-electroweak couplings are involved. It is performed by analysing \Ztautau decays in
which one $\tau$ decays leptonically ($\tau \rightarrow e/\mu+\nu\nu$) and the other hadronically ($\tau \rightarrow \text{hadron(s)}+\nu$). The leptonic
decay is utilised to trigger, select, and identify \Ztautau candidate events,
while the hadronic decay serves as a spin analyser.  
The $qq\rightarrow Z\rightarrow\tau\tau$ signal has been observed before by the ATLAS, CMS and LHCb collaborations~\cite{TOPQ-2012-05, ZtauCMS, ZtauLHCb}.
Due to the abundance
of background processes, strict requirements are applied to select a
sufficiently pure sample of \Ztautau decays from the proton--proton collision
data. Further requirements are dictated by the detector acceptance.  The
overall acceptance is larger for \Ztautau decays with left-handed $\tau^{-}$.
To provide a result that is close to the polarisation directly observed in the
selected signal region, the $\tau$ polarisation is measured in a fiducial region, which
is defined at stable-particle level and very similar to the selected signal region. The polarisation is predicted by using simulated event
samples produced with the \textsc{Alpgen}~\cite{ALPGEN} event generator interfaced with
the \textsc{Pythia6}~\cite{PYTHIA} parton shower and hadronisation model. The $\tau$ lepton
decay and spin effects are simulated with the \textsc{Tauola}~\cite{TAUOLA1} decay library using
$\sin^2\theta_\text{W}^\text{eff}=0.23147$ in the electroweak leading-order (LO) matrix element to
simulate polarisation and spin correlations in the
Tauola Universal Interface~\cite{TauUniInterface}. The prediction
in the fiducial region is $P_{\tau} = -0.270\pm0.006$.

The principal result presented in this paper is a measurement of the $\tau$
polarisation inside the \Zgamma mass range of 66~$< m_{\Zgamma} <$~116~GeV.
Away from the $Z$ boson mass peak, the degree of polarisation varies with
$m_{\Zgamma}$ and is determined by the interference between $Z$ boson- and photon-mediated
amplitudes. An inclusive measurement over a mass range around the $Z$ boson pole is
performed here, because the contributions slightly above and below the $Z$ boson pole
cannot be separated accurately. The $\Zgamma$ interference has approximately
the opposite effect on the polarisation below and above the $Z$ boson pole. Therefore,
and because the on-pole cross-section is dominant, the polarisation inside the
mass-selected region of 66~$< m_{\Zgamma} <$~116~GeV is close to $\Ptau$ at $\sqrt{s}=m_Z$.
The prediction by the \textsc{Alpgen} event generator interfaced with the \textsc{Pythia6} parton shower and hadronisation model and
\textsc{Tauola} library for $\tau$ decays is $\Ptau~=-0.1517\pm
0.0014\;(\text{stat})\pm0.0013\;(\text{syst})$. This is different from the $P_{\tau}$ value in fiducial region because some of the event selection requirements, such as transverse momenta thresholds, prefer one $\tau$ helicity state over another.
For the extrapolation from the selected signal region to the full
phase space inside the \Zgamma mass range, the $\tau\tau$ contribution is assumed to
originate from $\Ztautau$ decays. In particular, the spin correlations of the
two $\tau$ leptons are assumed to be those for unit-spin intermediate states.
The $\tau$ decays are assumed to follow the Standard Model expectations.
 
The first $\tau$ polarisation measurement at ATLAS was performed in $\Wtaunu$
decays in proton--proton collisions at the centre-of-mass energy of
$\sqrt{s}=7$~TeV recorded in 2010~\cite{STDM-2011-46}. 
The concept to extract the polarisation from a template fit to a polarisation sensitive observable is retained from that analysis.
To exploit the larger
dataset collected at $\sqrt{s}=8$~TeV, refined experimental techniques for
$\tau$ polarisation measurements at hadron colliders are utilised for
the measurement presented in this paper. In particular, the impact of
systematic uncertainties in the modelling of the polarisation observable for
signal events and the significant backgrounds are estimated
more thoroughly, because they are more important in the current measurement using a larger dataset. These techniques may serve as a foundation for future polarisation
measurements in decays of the Higgs boson or $\tau\tau$ final states with high invariant mass. A
good understanding of $\tau$ polarisation in $Z$ boson decays is indispensable
for these measurements. Moreover, the polarisation itself provides a potential
discriminant in Standard Model Higgs boson selection and searches for physics
beyond the Standard Model. In particular it may help to distinguish decays of
heavy particles where the same final states involving $\tau$ leptons are
predicted but with different helicity configurations, such as for separating $Z$ and $H$ or $A$
bosons or for distinguishing $W$ and $H^{\pm}$ bosons.

This paper is structured as follows. In Section~\ref{sec:detector} an overview
of the ATLAS detector is presented. The event samples, which were recorded by
ATLAS or simulated using the ATLAS simulation framework, are introduced in
Section~\ref{sec:samples}. The reconstruction and definition of physics objects
is documented in Section~\ref{sec:reco}. Section~\ref{sec:selection} describes
the selected signal region and the prediction of the polarisation in the fiducial region
and in the mass-selected region. The $\tau$ polarisation observable is introduced in
Section~\ref{sec:kinematics}. The estimation of the background contributions in
the selected signal region is documented in Section~\ref{sec:background}.
Section~\ref{sec:syst} describes the estimation of the experimental and theory
systematic uncertainties. A description of the fit model used to extract the
$\tau$ polarisation is given in Section~\ref{sec:fit}. The results of the
measurement are shown in Section~\ref{sec:result}, followed by conclusions in
Section~\ref{sec:conclusion}.

\section{ATLAS detector}
\label{sec:detector}

\newcommand{\AtlasCoordFootnote}{%
ATLAS uses a right-handed coordinate system with its origin at the nominal interaction point (IP)
in the centre of the detector and the $z$-axis along the beam pipe.
The $x$-axis points from the IP to the centre of the LHC ring,
and the $y$-axis points upwards.
Cylindrical coordinates $(r,\phi)$ are used in the transverse plane, 
$\phi\in(-\pi,\pi]$ being the azimuthal angle around the $z$-axis.
The pseudorapidity is defined in terms of the polar angle $\theta$ as $\eta = -\ln \tan(\theta/2)$.
Angular distance is defined as $\Delta R \equiv \sqrt{(\Delta\eta)^{2} + (\Delta\phi)^{2}}$. 
The $\Delta\phi$ separation is defined as $\mathrm{min}(|\phi_1-\phi_2|,2\pi-|\phi_1-\phi_2|)$.}

The ATLAS experiment~\cite{PERF-2007-01} at the LHC is a multi-purpose particle detector
with a forward-backward symmetric cylindrical geometry and a near $4\pi$ coverage in 
solid angle.\footnote{\AtlasCoordFootnote}
It consists of an inner tracking detector surrounded by a thin superconducting solenoid
providing a \SI{2}{\tesla} axial magnetic field, electromagnetic and hadronic calorimeters, and a muon spectrometer.
The inner tracking detector covers the pseudorapidity range $|\eta| < 2.5$.
It consists of silicon pixel, silicon microstrip, and transition radiation tracking detectors.
Lead/liquid-argon (LAr) sampling calorimeters provide electromagnetic (EM) energy measurements
with high granularity.
A hadronic (steel/scintillator-tile) calorimeter covers the central pseudorapidity range ($|\eta| < 1.7$).
The endcap and forward regions are instrumented with LAr calorimeters
for both the EM and hadronic energy measurements up to $|\eta| = 4.9$.
The muon spectrometer surrounds the calorimeters and features
three large air-core toroid superconducting magnets with eight coils each.
The field integral of the toroids ranges from \num{2.0} to \SI{6.0}{\tesla$\cdot$\metre} across most of the detector.
It includes a system of precision tracking chambers and fast detectors for triggering.
A three-level trigger system is used to select events~\cite{PERF-2011-02}.
The first-level trigger is implemented in hardware and uses a subset of the detector information
to reduce the accepted rate to at most \SI{75}{\kilo\hertz}.
This is followed by two software-based trigger levels that
together reduce the accepted event rate to \SI{400}{\hertz} on average
depending on the data-taking conditions during 2012.

\section{Data and simulated event samples}
\label{sec:samples}

The data sample was recorded by ATLAS in proton--proton collisions provided by the LHC at a centre-of-mass energy of $\sqrt{s}=8$~TeV in 2012. The integrated luminosity of the sample is $\mathcal{L}=20.2$~fb$^{-1}$ after beam and data quality requirements are satisfied. Candidate events are selected with four triggers, a single-muon or single-electron trigger requiring an isolated muon or electron with transverse momentum $\pt>24$~GeV complemented by higher-threshold ($\pt>60$~GeV for electrons, $\pt>36$~GeV for muons) triggers without isolation requirements. The accepted events must also contain at least one reconstructed primary vertex with more than three tracks with $\pt>400$~MeV each. If more than one such vertex is present, that with the highest sum of the squared transverse momenta of all associated tracks is chosen as the primary vertex.

The expected signal as well as several background processes are modelled using
samples of simulated events. Signal (\Ztautau) + jets events were generated with
boson masses $m_{\Zgamma}>60$~GeV with the \textsc{Alpgen} event generator interfaced with
the \textsc{Pythia6} fragmentation, hadronisation and underlying event (UE)
modelling. The \textsc{Alpgen} event generator was used with default electroweak parameters~\cite{ALPGEN}. The
\textsc{CTEQ6L1}~\cite{CTEQ6L1} parton distribution function (PDF) set and a set of tuned parameters called the
Perugia2011C tune~\cite{Perugia} were used. QED radiation was simulated by the \textsc{Photos}~\cite{PHOTOS} algorithm.
The information about the $\tau$ helicity state was not stored at the generation step for the (\Ztautau) + jets process.
The spin polarisation and correlations were therefore simulated using Tauola Universal Interface~\cite{TauUniInterface} as expected
from the electroweak lowest-order matrix element for the \Ztautau production process, with $\sin^2\theta_\text{W}^\text{eff}=0.23147$. The $\tau$ decays were simulated using the \textsc{Tauola} decay library~\cite{TAUOLA1}.
The helicities of $\tau$ leptons
generated by the \textsc{Tauola} algorithm were not stored so that the
helicity is reconstructed in the generated signal samples with the
\textsc{TauSpinner}~\cite{TS_czyczula} package associated with
the \textsc{Tauola} decay library. The \textsc{TauSpinner} algorithm assigns the helicity of
$\tau$ leptons randomly based on probabilities derived from the kinematic configuration of the $\tau$ decays. The signal sample is thereby split into events with left-handed
$\tau^{-}$ (and right-handed $\tau^{+}$) and those with right-handed $\tau^{-}$
(and left-handed $\tau^{+}$). The \textsc{TauSpinner} algorithm averages over incoming parton flavours and four-momenta
whereas the \textsc{Tauola} algorithm directly accesses the incoming partons in each event.
The average over initial parton states is performed using the MRSTMCal PDF set~\cite{MRSTMcal} 
in this analysis. Spin correlations as expected in $\Ztautau$ decays are assumed.
The \textsc{TauSpinner} package was extensively tested and validated by its authors~\cite{TS_acta, TS_czyczula, Richter-Was2016}
and used in several measurements~\cite{EXOT-2014-05, CMS_TS}.

For studies of systematic uncertainties, an auxiliary sample of \Ztautau
events was produced using the \textsc{Pythia8}~\cite{PYTHIA8} event generator with the \textsc{CTEQ6L1} PDF set and
AU2~\cite{ATL-PHYS-PUB-2012-003} tune for the UE. In this case the
\textsc{Pythia8} event generator was used to model both the production process and decays including
those of $\tau$ leptons. Further auxiliary signal samples were
produced with the \textsc{Powheg}~\cite{POWHEG1, POWHEG2, POWHEG3} event generator 
interfaced with the \textsc{Pythia8} parton shower simulation using the \textsc{CT10} PDF set~\cite{CT10} 
and with the \textsc{Alpgen} event generator interfaced with the \textsc{Herwig/Jimmy}~\cite{HERWIG65,Jimmy} hadronisation and UE modelling.
Only stable-particle-level information is used in the auxiliary samples.

Background samples of simulated $(\Wenu)$ + jets, $(\Wmunu)$ + jets, $(\Wtaunu)$ + jets, $(\Zee)$ + jets, and
$(\Zmumu)$ + jets events were generated using the \textsc{Alpgen} event generator interfaced with the \textsc{Pythia6} hadronisation modelling
and with the same settings as for the signal sample. For these samples, LO matrix elements were calculated
for up to five additional partons. The resulting predictions were scaled such that the total cross-sections match the respective inclusive
next-to-next-to-leading-order (NNLO) predictions~\cite{WZNNLO}. A sample of top 
pair production was generated using the \textsc{Powheg}~\cite{POWHEG1, POWHEG2,
POWHEG3} event generator interfaced with the \textsc{Pythia6} hadronisation modelling and with the CT10~\cite{CT10} PDF
set. The $\ttbar$ cross-section was calculated at NNLO+NNLL (next-to-next-to-leading-logarithm)~\cite{POWHEG2}.
In this analysis all simulated event samples receive data-driven corrections to the normalisation predicted 
by the aforementioned cross-sections with the exception of the $\ttbar$ background.
The list of simulated event samples used in this analysis is given in
Table~\ref{tab:samples}.

\begin{table}[tb]
\begin{center}
\begin{small}
\begin{tabular}{llll}
\toprule
Sample              & Event generator                & PDF & UE tune \\
\midrule
$(\Ztautau)$ + jets & \textsc{Alpgen 2.14} \cite{ALPGEN} + \textsc{Pythia6.427} \cite{PYTHIA}      & \textsc{CTEQ6L1} \cite{CTEQ6L1} & Perugia2011C  \cite{Perugia}  \\
$(\Ztautau)$ + jets & \textsc{Pythia 8.160} \cite{PYTHIA8}                                     & \textsc{CTEQ6L1}                & AU2 \cite{ATL-PHYS-PUB-2012-003} \\
$(\Ztautau)$ + jets & \textsc{Powheg} r1556 \cite{POWHEG1, POWHEG2, POWHEG3} + \textsc{Pythia 8.160} & CT10~\cite{CT10}                & AUET2~\cite{ATL-PHYS-PUB-2011-008} \\
$(\Ztautau)$ + jets & \textsc{Alpgen 2.14} + \textsc{Herwig 6.5/Jimmy 4.3} \cite{HERWIG65,Jimmy}                 & \textsc{CTEQ6L1}                & Perugia2011C   \\
\midrule
Top pairs + jets    & \textsc{Powheg} r2129 + \textsc{Pythia 6.426} & CT10 & AUET2 \\
\midrule
$(\Wenu)$   + jets  & \textsc{Alpgen 2.14} + \textsc{Pythia 6.427} & \textsc{CTEQ6L1} & Perugia2011C\\
$(\Wmunu)$ + jets   & \textsc{Alpgen 2.14} + \textsc{Pythia 6.427} & \textsc{CTEQ6L1} & Perugia2011C\\
$(\Wtaunu)$ + jets  & \textsc{Alpgen 2.14} + \textsc{Pythia 6.427} & \textsc{CTEQ6L1} & Perugia2011C\\
\midrule                                
$(\Zee)$ + jets     & \textsc{Alpgen 2.14} + \textsc{Pythia 6.427} & \textsc{CTEQ6L1} & Perugia2011C\\
$(\Zmumu)$ + jets   & \textsc{Alpgen 2.14} + \textsc{Pythia 6.427} & \textsc{CTEQ6L1} & Perugia2011C\\
\bottomrule
\end{tabular}
\end{small}
\end{center}

\caption{
  Simulated event samples used in the analysis. The table lists the sample, the event generator, the PDF set, and the underlying-event tune. \label{tab:samples}
}
\end{table}

The simulated $\Zgamma$ boson decays (in both the signal and background
processes) are reweighted such that the simulated $\pt$ spectrum of the
$\Zgamma$ bosons matches the observed $\pt$ spectrum in data, as done in
Ref.~\cite{PERF-2014-06}, using $\Zmumu$ events.
The response of the ATLAS detector was simulated~\cite{SOFT-2010-01} using
\textsc{Geant4}~\cite{geant4}. Simulated events were overlaid with additional
minimum-bias events generated with the \textsc{Pythia8} event generator to account for the effect
of multiple interactions occurring in the same and neighbouring bunch crossings
(pile-up). The simulated events were re-weighted such that the distribution of
the average number of pile-up interactions per bunch crossing matches the observed spectrum in data.  
Finally, the simulated events were processed through the same
reconstruction algorithms as the data.

\section{Event reconstruction and object definitions}
\label{sec:reco}

Electrons are reconstructed from energy clusters in the calorimeter which have a matching track in the inner detector. Electron candidates are considered if they satisfy \enquote{loose} identification criteria~\cite{ATLAS-CONF-2014-032} and the requirements of $\pt>15$~GeV and $|\eta| < 2.47$. 

Muon candidates are reconstructed from associated tracks in the inner detector and the muon spectrometer. They are required to satisfy \enquote{loose}~\cite{PERF-2014-05} identification criteria as well as the requirements of $\pt>10$~GeV and $|\eta| < 2.5$. The electron and muon (lepton) candidates that pass the aforementioned requirements are in the following referred to as preselected.

In order to be selected, lepton candidates are required to have $p_\text{T,lepton}>26$~GeV
and to pass stricter identification requirements. Specifically, electron
candidates must satisfy \enquote{tight}~\cite{ATLAS-CONF-2014-032}
identification criteria and lie outside the calorimeter transition region of
$1.37<|\eta|<1.52$. Muon candidates are required to have a combined track~\cite{PERF-2014-05} in the
inner detector and muon spectrometer.  Additionally, isolation requirements in
the inner detector and calorimeter are applied to both the electrons and muons.
The fraction of the momentum carried by tracks other than the identified lepton
track inside a cone of size $\Delta R=0.4$ around the lepton track must be less than
6\%. Similarly, after correcting for pile-up, the fraction of the transverse energy
reconstructed in a cone of size $\Delta R=0.2$ around the lepton axis but not
associated with the lepton candidate must not exceed 6\% of the lepton's transverse
energy.

Jets are reconstructed~\cite{Cacciari:2011ma} with the anti-$k_{t}$ algorithm~\cite{antikt} with a radius parameter $R=0.4$ using topological clusters of energy deposits in the electromagnetic and hadronic calorimeters within $|\eta|<4.5$ with a local hadronic calibration~\cite{PERF-2014-07}. In this analysis, jets with $p_{\mathrm{T}} > 20$~GeV are used in the calculation of missing transverse momentum. Here,
jets with $|\eta|<2.4$ and $\pt<50$~GeV must meet additional criteria designed to select jets from the hard-scatter interaction and reject those originating from pile-up:
among the tracks associated with the jet, those originating from the primary vertex must contribute at least 50\% of the sum of the scalar \pt of all those tracks~\cite{TOPQ-2013-12}.
In this analysis, no selection is made on the number of jets.

The reconstruction of $\tau$ candidates is based on the visible decay
products of hadronically decaying $\tau$ leptons ($\tauhad$ with visible
component $\tauhadvis$). These candidates are seeded by jets reconstructed
with transverse momentum above 10~GeV.
At this stage of the analysis $\tauhad$ candidates are required to have reconstructed $p_{\text{T},\tauhadvis}>20$~GeV and
$|\eta|<2.47$, to have exactly one or three charged-particle tracks, to
be identified with \enquote{medium} identification criteria~\cite{taureco}, and
to have reconstructed electric charge of $\pm 1$. The $\tauhad$ energy scale is determined
from simulated event samples and accounts for the mixture of hadrons typical of
$\tauhad$ decays as well as contributions from the UE, pile-up,
and energy outside of the $\tauhadvis$ cone~\cite{taureco}. A \enquote{medium}
electron veto as well as a muon veto are applied to reject electrons and muons
that are reconstructed as $\tauhad$ candidates~\cite{taureco}.

Objects that are reconstructed in geometrically overlapping regions, given by a cone of size $\Delta R=0.2$, are identified with the above definitions with the following precedence: preselected muon, preselected electron, $\tauhad$ candidate, and jet. For the purpose of removing overlaps between muons and $\tauhad$ candidates, the $\pT$ threshold for muon candidates is reduced to 2~GeV.

The missing transverse momentum $(\met)$ is calculated as the modulus of the negative vectorial sum of the \textbf{$\mathrm{p_T}$} of all fully reconstructed and calibrated physics objects in the event, as well as a term for the remaining activity in the calorimeter~\cite{PERF-2011-07}. Here, preselected leptons are included in the sum.

\section{Event selection}
\label{sec:selection}

Selection criteria are applied to obtain a sample enhanced in $\Ztautau$ events where one of the $\tau$ leptons decays leptonically ($\taulep$) and the other hadronically. 
The \tauhad candidate is required to have exactly one charged-particle track (single-prong).
Events are categorised into channels by the lepton flavour (electron or muon), which are referred to as $\tau_e\text{--}\tauhad$ and $\tau_\mu\text{--}\tauhad$ channels. The kinematic requirements on electrons and muons are similar and, therefore, the event selections that define the two selected signal regions are described in parallel.

Exactly one $\tauhad$ candidate and exactly one lepton that fulfil the respective selection criteria and that have opposite-sign electric charges are required. 
Two selection requirements are implemented to reduce the significant background that arises from $W$+jets production in which a lepton is 
reconstructed correctly and a jet is misidentified as a $\tauhad$ candidate. The transverse mass, $m_{\mathrm{T}}$, built from the lepton and missing transverse momenta, is defined as
\begin{equation}
 m_{\mathrm{T}} = \sqrt{2~p_{\mathrm{T,lepton}}~\met~\left(1-\mathrm{cos}\left(\Delta\phi\left(\mathrm{lepton},\met\right)\right)\right)} \nonumber
\end{equation}
and is required to satisfy $m_{\mathrm{T}}<30$~GeV. The sum of the azimuthal angular separation between the $\tauhad$ candidate and the $\met$ directions, and the lepton and the $\met$ directions, 
\begin{equation}
\sum\Delta\phi = \Delta\phi\left(\tauhadvis,\met\right)+\Delta\phi\left(\mathrm{lepton},\met\right) \nonumber
\end{equation}
is required to satisfy $\sum\Delta\phi < 3.5$. This requirement suppresses event topologies in which the \met lies outside of the angle spanned by the $\tau$ candidate and the lepton, which are common for $W$+jets processes and rare for signal events.
In addition, the visible mass of the $\tauhad$ candidate and lepton, $m_\mathrm{vis} = m(\tauhadvis,\text{lepton})$, is required to satisfy 40~$< m_{\mathrm{vis}}<85$~GeV to further reduce backgrounds, notably the non-signal $\Zgamma$+jets background in which the $\Zgamma$ boson decays to electron or muon pairs. For signal events around the $Z$ boson pole that pass the previous requirements, the $m_\mathrm{vis}$ distribution is centred at about $66~$GeV and has a width of about $10~$GeV. This is insufficient for separating $\Ztautau$ decays on and off the $Z$ boson pole. The selection criteria described above define the selected signal region of this analysis.

Some of the object and event selection requirements have different acceptances
for signal decays with one specific~$\tau^{-}$ helicity state: the
$p_{\mathrm{T,lepton}}$ requirement is about twice as efficient for $\Ztautau$
events with leptonically decaying left-handed $\tau^{-}$ leptons as for those with leptonically decaying right-handed
$\tau^{-}$ leptons. Here, the polarisation of the \tauhad is affected due to
spin correlations resulting from angular momentum conservation in \Ztautau
decays. This is partially counteracted by the $p_{\mathrm{T},\tauhadvis}$ and
$m_{\mathrm{T}}$ requirements. These biases result from dependencies of
the $\tau$ lepton momentum share carried by neutrinos on the helicity state and
the respective decay modes. The size of this effect may be different for possible unexpected contributions from
physics processes other than from intermediate states with unit spin decaying to $\tau$ pairs.
Hence the polarisation is also measured in a fiducial region which is defined with stable-particle-level quantities (see Table~\ref{tab:fidregion_def}).
It corresponds very closely to the selected signal region. For the extraction of
the $\tau$ polarisation in this region, the simulated signal sample is split into
three components:
\begin{itemize}
 \item Events inside the fiducial region with left-handed $\tau^{-}$ leptons,
 \item Events inside the fiducial region with right-handed $\tau^{-}$ leptons,
 \item Events outside the fiducial region.
\end{itemize}
About 80\% of the events in the selected signal region originate from the fiducial region. Most of the remaining events fail the $m_\mathrm{T}$ , $p_{\mathrm{T},\tauhadvis}$ or $p_{\mathrm{T,lepton}}$ requirements on stable-particle level but pass them at reconstructed-detector level.

\begin{table}[tb]
  \begin{center}
   \begin{tabular}{ll}
    \toprule
    One $\taulep$ decay                                                     & One single-prong $\tauhad$ decay   \\
    $p_{\mathrm{T,lepton}}>26$~GeV                                          & $p_{\mathrm{T},\tauhadvis}>20$~GeV \\
    $|\eta_{e}|<2.47$ and not $1.37<|\eta_{e}|<1.52$  or $|\eta_\mu| < 2.5$ & $|\eta_{\tauhadvis}|<2.47$      \\
    \midrule
    $m_{\mathrm{T}}<30$~GeV                                                 & 40~$< m_{\mathrm{vis}}<85$~GeV  \\
    \bottomrule
   \end{tabular}
  \end{center}
 \caption{
  Definition of fiducial region for $\Ztautau$ decays. The requirements are applied at stable-particle level. Here, the \met is calculated from the momenta of the neutrinos that originate from the $\Ztautau$ decays.
  \label{tab:fidregion_def} 
 }
\end{table}

For the extraction of the $\tau$ polarisation in $\Ztautau$ decays inside the
mass-selected region of 66~$<m_{\Zgamma} <$~116~GeV, the signal sample is split into these
components:
\begin{itemize}
 \item Events with $m_{Z/\gamma^{*}}$ inside the mass-selected region with left-handed $\tau^{-}$ leptons,
 \item Events with $m_{Z/\gamma^{*}}$ inside the mass-selected region with right-handed $\tau^{-}$ leptons,
 \item Events with $m_{Z/\gamma^{*}}$ outside the mass-selected region,
\end{itemize}
where the mass-selected region is defined at stable-particle level. About $98$\% of the
simulated $\Ztautau$ events in the selected signal region originate from the mass-selected region.

The $\tau$ polarisation is measured using the $\tauhad$ decay as a spin
analyser, and without utilising spin correlations of the two $\tau$ leptons. Therefore, the
polarisation measurement in the fiducial region does not strongly rely on the
prediction of the $\tau$ spin correlations. The most important exception is
that the contribution of $\Ztautau$ events which are outside the fiducial
region but which fall inside the selected signal region is taken from simulation. In contrast, the
polarisation measurement in the mass-selected region relies on the prediction of the spin
correlations when extrapolating to the full phase space and is therefore more model-dependent. Because of that, 
the interpretation of the measurement in the mass-selected region is largely model-dependent, if an anomalous polarisation value is measured.

The theoretical prediction of the $\tau$ polarisation in the mass-selected region of 66~$<
m_{\Zgamma} <$~116~GeV is obtained by performing a fit to the distribution of
the momentum fraction, $x$, carried by the $\pi^{\pm}$ at stable-particle level in
$\tau^\pm\to\pi^\pm\nu$ decays for events inside the mass-selected region. Specifically, this
distribution follows $f(x) = 1 + P_{\tau}(2x-1)$ as described in
Ref.~\cite{pienergyspect}. The resulting prediction is $\Ptau~=-0.1517\pm
0.0014\;(\text{stat})\pm0.0013\;(\text{syst})$. It is unaffected by
\textsc{TauSpinner} and MC-related systematic uncertainties and the quoted
uncertainty results from the choice of shower model simulation and PDFs. Since
the $x$ distribution is altered by the fiducial region selection, the
polarisation in the fiducial region can only be predicted from the numbers of
events in which the $\tau^{-}$ is classified as left- and right-handed by
\textsc{TauSpinner}. This method is affected by
\textsc{TauSpinner} systematic uncertainties, so the prediction of the
polarisation in the fiducial region is less accurate than that of the
polarisation in the mass-selected region. A predicted value of $P_{\tau} =-0.270\pm0.006$ is
obtained. Details of the estimation of particular systematic uncertainties are
given in Section \ref{sec:theounc}.

\section{Observable for $\tau$ polarisation}
\label{sec:kinematics}

The helicity of the $\tau$ lepton 
manifests itself in the kinematic distributions of its decay products.

The $\tau$ decay mode exhibiting the highest sensitivity to the $\tau$ polarisation is $\tau^{\pm} \to h^{\pm} \nu$, where $h^{\pm}$ denotes $\pi^{\pm}$ or $K^{\pm}$  (branching ratio, $\mathcal{B}$, $\simeq11.5\%$~\cite{Olive:2016xmw}).
The branching ratio of the decay mode involving a $\pi^{\pm}$ exceeds that of the mode involving a $K^{\pm}$ by more than an order of magnitude.
This also holds for the $\tau$ decay modes described below. 
In the $\tau$ rest frame, the neutrino (always left-handed) is preferentially
emitted opposite to the $\tau^{-}$ spin orientation. 

The angle $\theta$ between the $\tau$ flight direction in the laboratory frame and $\pi^{\pm}$ flight direction in the $\tau$ rest frame is the primary observable sensitive to $\tau$ polarisation. 
It cannot be measured directly at hadron colliders because insufficient information about the initial state is available. However, $\theta$ affects the momentum fraction carried by the $h^{\pm}$ resulting in a larger acceptance for right-handed than for left-handed $\tau^{-}$ in $\tau^{-}\to h^{-} \nu$ decays. 

Another $\tau$ decay mode, $\tau^{\pm} \to h^{\pm } \pi^0 \nu$ ($\mathcal{B}\simeq25.9\%$~\cite{Olive:2016xmw}), plays an important role in the polarisation measurement. It offers the kinematic simplicity of a two-body decay, since it goes mostly through sequential decays $\tau^{\pm} \to \rho^{\pm }  \nu$, $\rho^{\pm } \to \pi^{\pm} \pi^0$,  but 
the sensitivity to the angle between the $\tau$  direction of flight and
$\pi^{\pm}$ is lower, due to the mixing  
of longitudinally and transversely polarised $\rho^{\pm }$ vector mesons. 
The products of the $\rho^{\pm } \to \pi^{\pm} \pi^0$ decay are experimentally accessible and their angular distributions as well as their energies depend on the helicity of the vector meson.

The angle between
the direction of flight of the $\rho^{\pm }$ meson and $\pi^{\pm}$ in the $\rho^{\pm }$ rest frame is related to the energy--sharing between the $\pi^{\pm}$ and the $\pi^0$ and is sensitive to the $\tau$ helicity.
An asymmetry of energies carried by the charged and neutral pions and measured in the laboratory frame is defined as:
 \begin{equation}
 \Upsilon_{\mathrm{theory}} = \frac{E_{\pi^{\pm}} - E_{\pi^0}}{E_{\pi^{\pm}} + E_{\pi^0}}.
 \label{eq:obs_th}
 \end{equation}
This asymmetry  carries high sensitivity to polarisation and was effective in measuring the $\tau$ polarisation in the decay $W\to \tau \nu$~\cite{STDM-2011-46}.

The other decay modes considered are the modes with more neutral pions ($\tau^{\pm} \to h^{\pm } N\pi^0 \nu$, $N\ge 2$), and decay modes with three charged mesons, where two tracks are lost, and a small admixture of other modes. 
In this class of decay modes the dominant mode is $\tau^{\pm} \to h^{\pm } 2\pi^0 \nu$, with $\mathcal{B}\simeq9.3\%$~\cite{Olive:2016xmw}. 
It has more complicated kinematics than $\tau^{\pm} \to h^{\pm } \pi^0 \nu$, but it nonetheless contributes to the polarisation sensitivity. 
The contributions from other channels are small. For example the branching ratio of $\tau^{\pm} \to h^{\pm } 3\pi^0 \nu$ is only $\simeq1\%$~\cite{Olive:2016xmw}.

The asymmetry defined in Eq.~(\ref{eq:obs_th}) is approximated using the experimental observables.
In this approach 
the $\pt$ of a single track associated with the $\tauhad$~candidate replaces the energy of the $\pi^{\pm}$. Since the energies of neutral pions are not measured directly, 
the difference between the $\tau$~lepton visible $E_\text{T}$, defined below, and the track~$\pt$ is used in place of the $\pi^0$ energy. 
  As the minimum $\tauhad$ $\pt$ required is 20~GeV, the $\tau$ leptons are relativistic enough to use this approximation. The visible $E_\text{T}$ of $\tauhad$ candidates is reconstructed using the energy deposit in the calorimeter~\cite{taureco}.
Therefore, the charged asymmetry is given by:

\begin{equation}
    \Upsilon = \frac{\et^{\pi^{\pm}} - \et^{h^0}}{\et^{\tauhadvis}} = 2\frac{\pt^{\mathrm{track}}}{\et^{\tauhadvis}} - 1,
    \label{eq:obs_exp}
\end{equation}
where $h^0$ denotes neutral particles produced in the $\tau$ decay, which are
mostly neutral pions.

\begin{figure}[htbp]
    \begin{center}
        {
      \includegraphics[width=0.46\textwidth,angle=0]{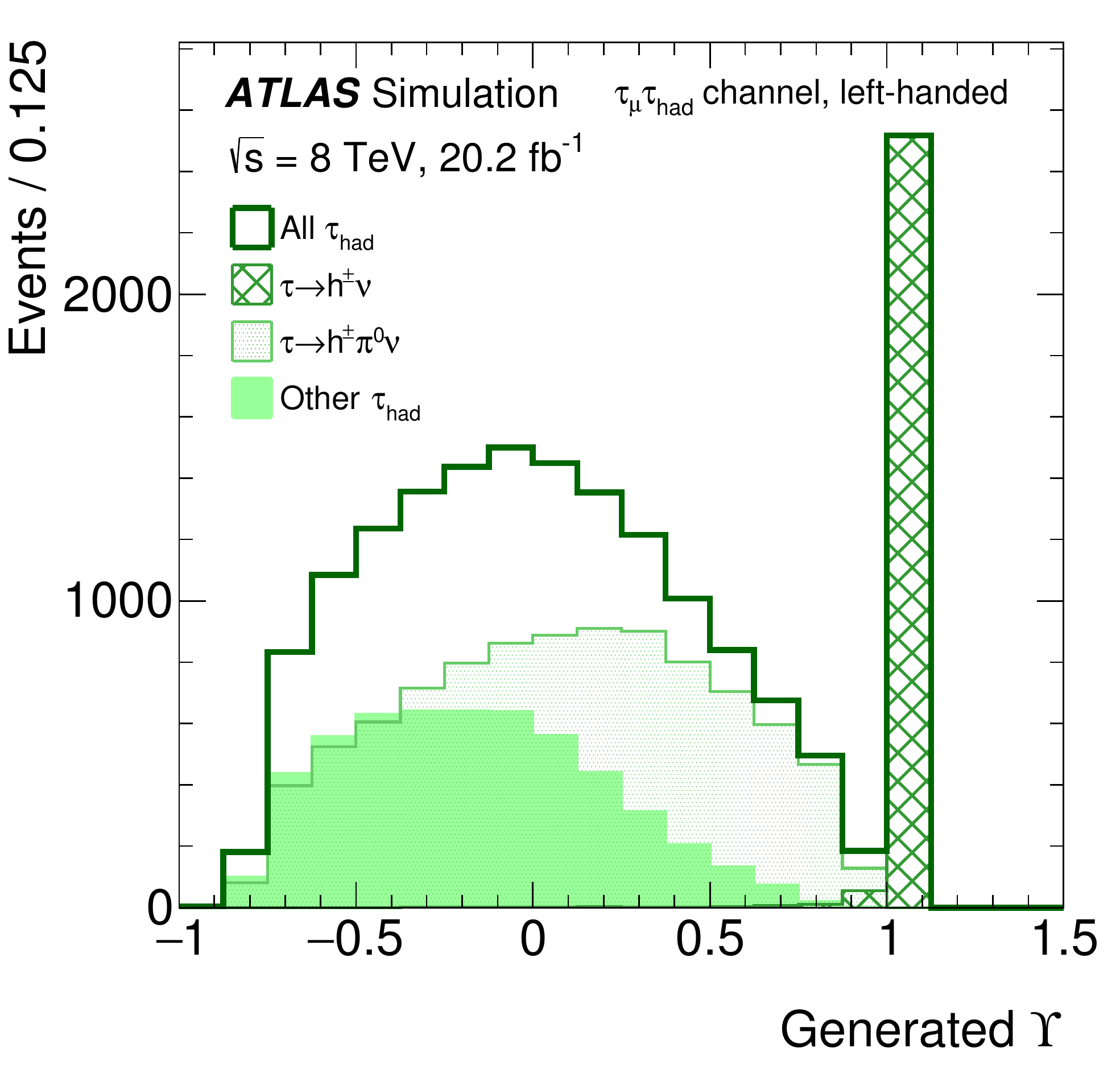}
      \includegraphics[width=0.46\textwidth,angle=0]{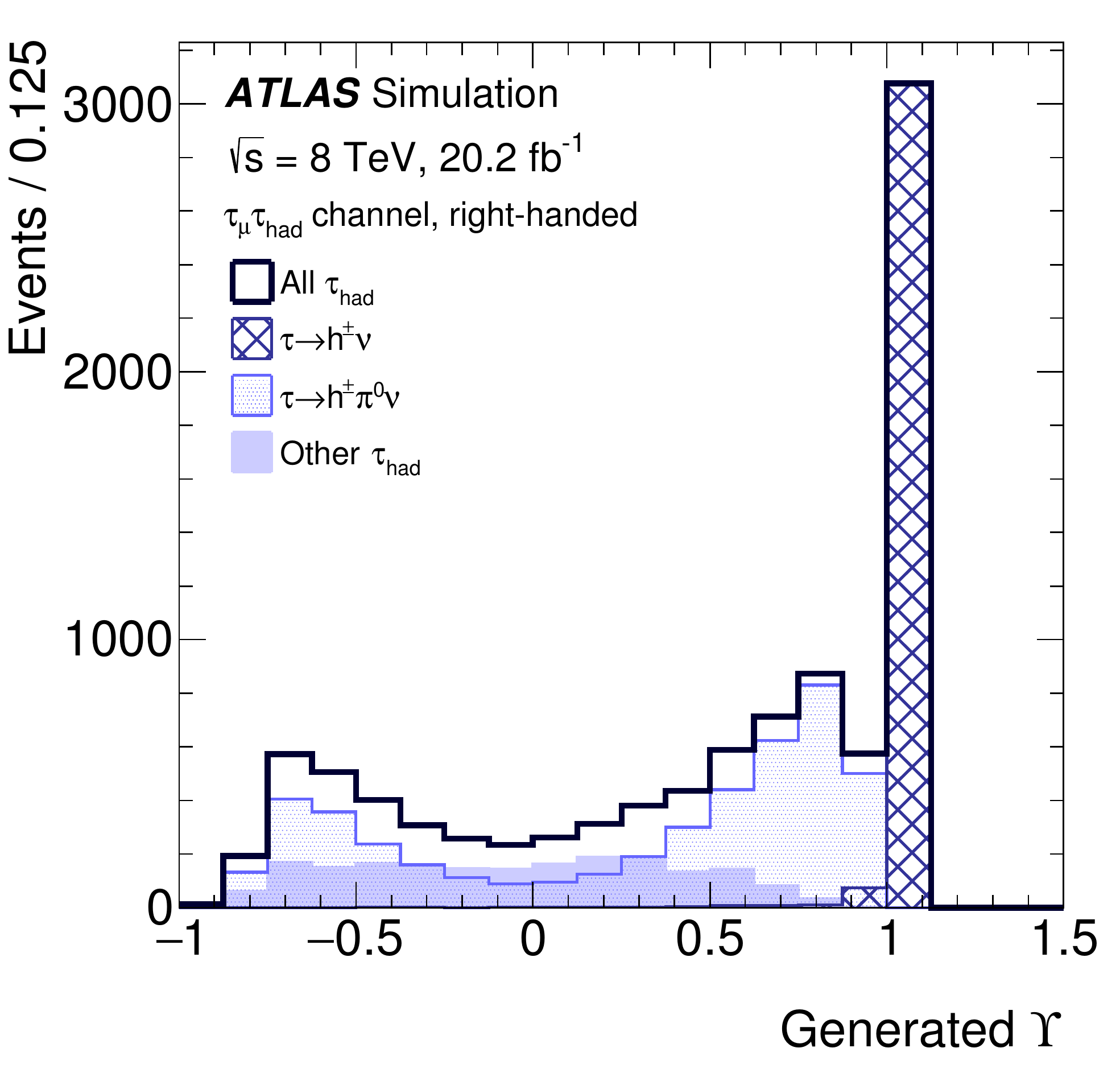}
      \includegraphics[width=0.46\textwidth,angle=0]{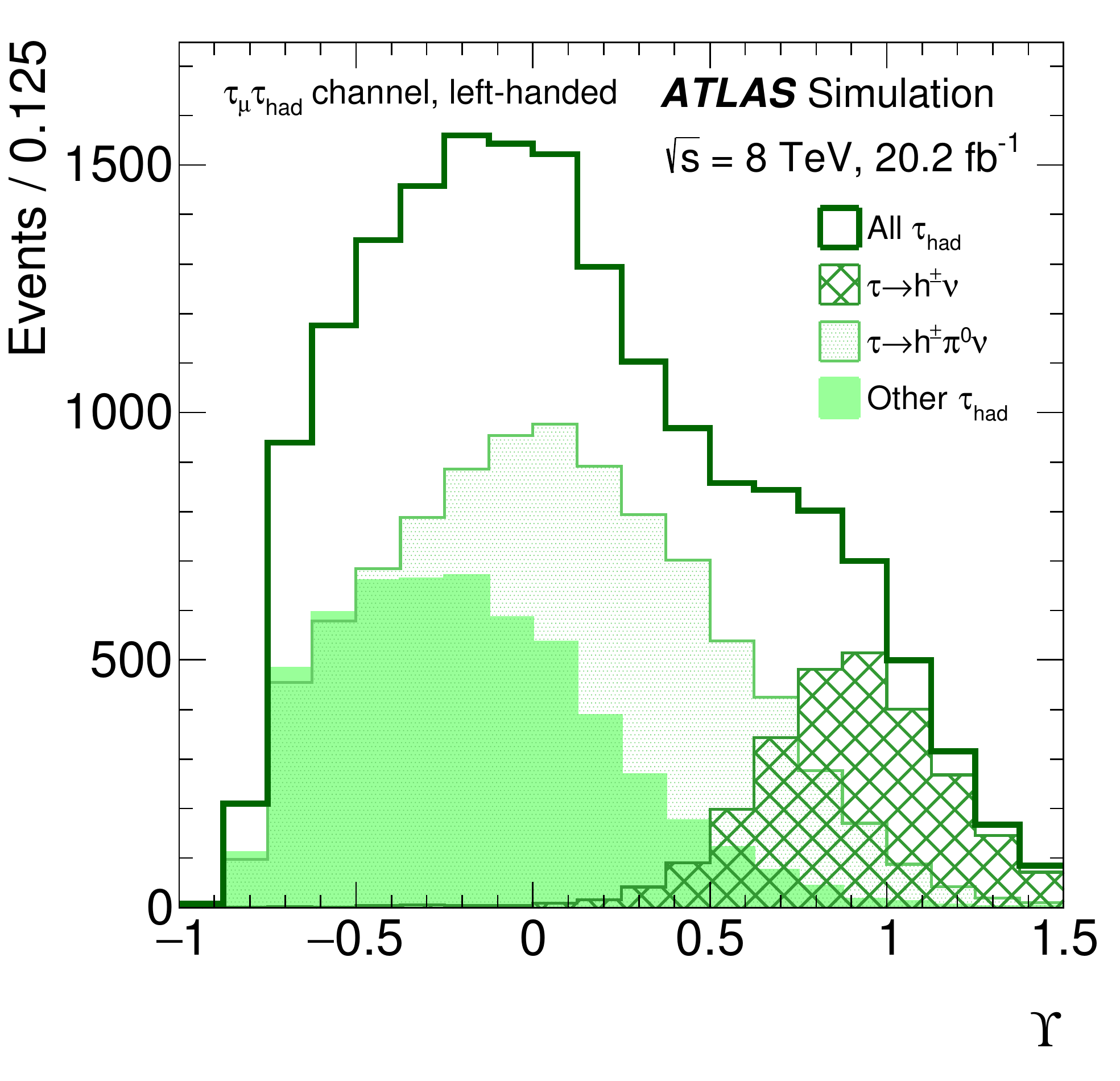}
      \includegraphics[width=0.46\textwidth,angle=0]{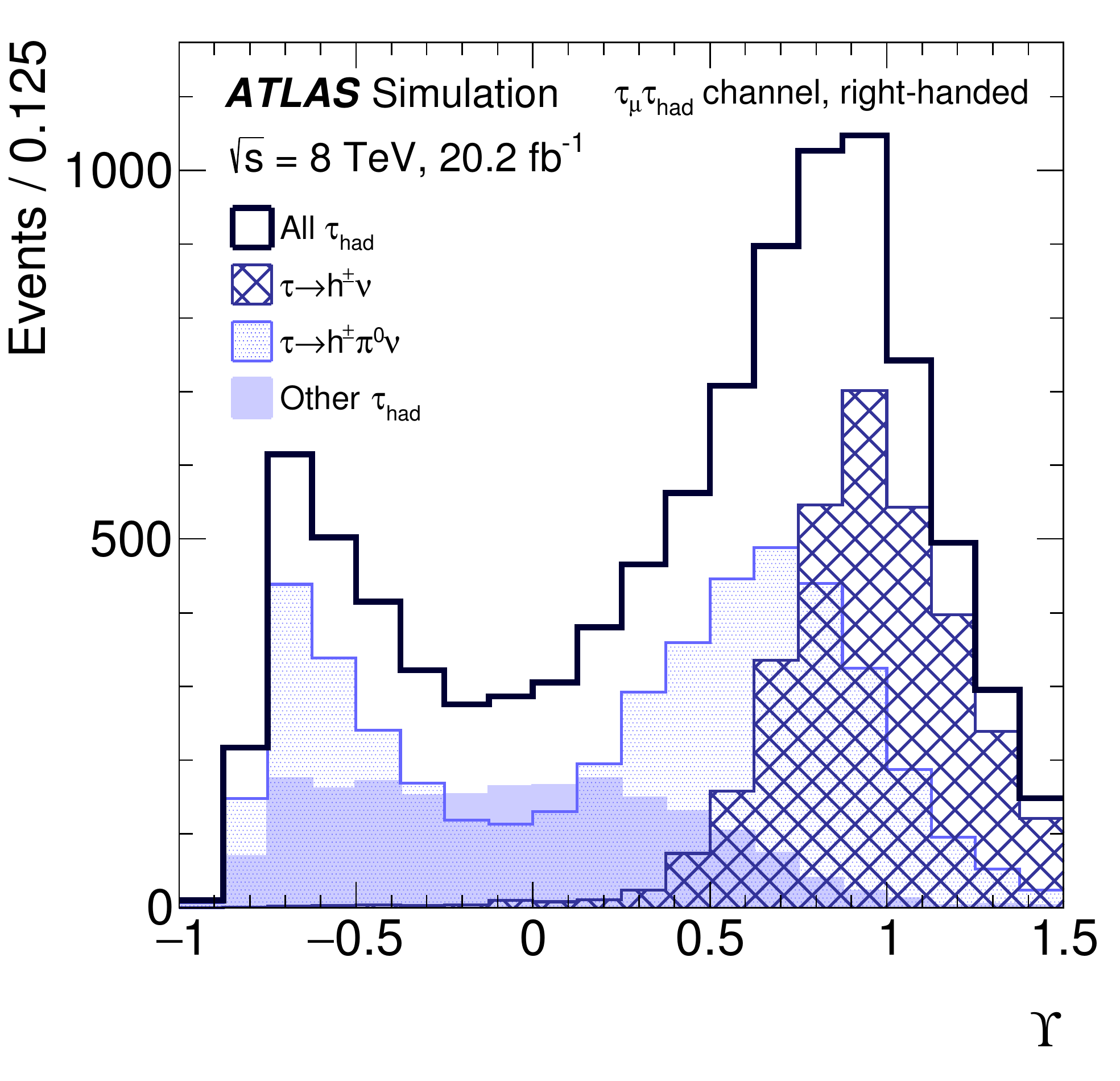}
        }
    \end{center}
    \caption[]{
    Charged asymmetry distributions as defined in Eq.~(\ref{eq:obs_exp}) for
    left-handed (left) and right-handed (right) single-prong reconstructed
    $\tauhad$ leptons in simulated \Ztautau  decays after the full event selection
    in the $\tau_\mu\text{--}\tauhad$ channel. The charged asymmetry is calculated from
    stable-particle-level (top) and reconstructed-detector-level (bottom) quantities. In addition to the
    inclusive distributions, the constituent distributions corresponding to
    generated $\tau$ leptons that decay in the $\tau\rightarrow
    h^{\pm}\nu$ and $\tau\rightarrow h^{\pm}\pi^0\nu$ ($h^{\pm}$
    denotes $\pi^{\pm}$ or $K^{\pm}$) modes are overlaid, as well as that of
    the remaining decay modes. The latter mainly consist of $\tau\rightarrow
    h^{\pm}N\pi^0\nu$ decays, where $N \ge 2$. The analysis does not, 
    however, distinguish between the decay modes.
    The distributions are
    normalised according to their respective cross-sections. Here, the polarisation is
    taken from the simulation.
    \label{fig:upsilon_sim}
  }
\end{figure}

The shapes of $\Upsilon$ distributions for the left-handed and right-handed reconstructed single-prong $\tau$ candidates obtained from simulation after the full event selection are shown in Figure~\ref{fig:upsilon_sim}. 
 
The $\Upsilon$ spectra include effects that originate from the acceptance, object reconstruction, and efficiencies as well as the event selection. 

The $\Upsilon$ distributions for left- and
right-handed $\tau$ leptons have different shapes in case of the
$\tau^{\pm}\rightarrow h^{\pm} N \pi^0\nu$, $N\ge1$ decay modes,
 while for the $\tau^{\pm}\rightarrow h^{\pm} \nu$ mode the polarisation sensitivity comes mostly from 
different acceptances and efficiencies. The branching ratio of the
$\tau^{\pm}\rightarrow h^{\pm}\pi^0\nu$ decay mode also exceeds the total
branching ratio of the remaining single-prong $\tauhad$ decay modes combined. 

Most of the sensitivity originates from the $\tau^{\pm}\rightarrow
h^{\pm}\pi^0\nu$ decays. The $\tau^{\pm}\rightarrow h^{\pm}\nu$ and
other remaining modes have similar individual sensitivities and they also make
a significant contribution to the overall polarisation sensitivity.

\section{Background estimate}
\label{sec:background}

The signal topology can be mimicked by several background processes, which
require different strategies for their estimation. The two largest background
contributions arise from multijet and $W$+jets events. In multijet events both
the lepton and \tauhad candidates originate from quark- or gluon-initiated
jets. They contribute about 19\% (7\%) of the total event yield in the
$\tau_e\text{--}\tauhad$ ($\tau_\mu\text{--}\tauhad$) channel. In most of the $W$+jets
background events, a lepton is produced in the decay of the $W$ boson and a jet
is misidentified as a \tauhad lepton. They contribute about 7\% (8\%) of the events
in the $\tau_e\text{--}\tauhad$ ($\tau_\mu\text{--}\tauhad$) channel. Both major backgrounds are estimated using data-driven
techniques, which are described in this section. The control regions utilised
for these estimates are compiled in Table~\ref{tab:ControlRegionsOverview}.  
A minor background contribution consists of $(\Zll)$+jets ($\ell=e, \mu$)
events, where \tauhad candidates can originate from quark- or gluon-initiated
jets or from one of the leptons. Another background stems from events with top
pairs which involve a real lepton and either a real \tauhad or a quark- or
gluon-initiated jet that is misidentified. These minor background contributions
are estimated from the simulation. They are normalised with their respective
cross-sections and corrections for differences in (mis-) identification between
data and the simulation are applied. They amount to about 5\% (2\%) of the
total event yield in the $\tau_e\text{--}\tauhad$ ($\tau_\mu\text{--}\tauhad$)
channel.

\begin{table}[tb]
 \centering
  \begin{small}
   \begin{tabular}{ll}
    \toprule
    Region                                & Event selection changes compared to selected signal region \\
    \midrule
    Same-sign region                      & inverted opposite-charge-sign requirement \\ \addlinespace
    Opposite-sign multijet control region & inverted lepton-isolation requirement \\ \addlinespace
    Same-sign multijet control region     & inverted lepton-isolation and opposite-charge-sign requirements \\ \addlinespace
    Opposite-sign $W$+jets control region & $\sum\Delta\phi\geq3.5$, $m_{\text{T}} > 70$~GeV (instead of $\sum\Delta\phi < 3.5$, $m_{\text{T}} < 30$~GeV) \\ \addlinespace
    Same-sign $W$+jets control region     & $\sum\Delta\phi\geq3.5$, $m_{\text{T}} > 70$~GeV (instead of $\sum\Delta\phi < 3.5$, $m_{\text{T}} < 30$~GeV), \\
                                          & inverted opposite-charge-sign requirement \\

    \bottomrule
   \end{tabular}
 \end{small}
 \caption{Summary of the control regions used for the background estimates.}
 \label{tab:ControlRegionsOverview}
\end{table}

\subsection{Estimation of $W$+jets background}
\label{sec:Wjets}

\begin{figure}[htbp]
  \begin{center}
{
   \includegraphics[width=0.48\textwidth,angle=0]{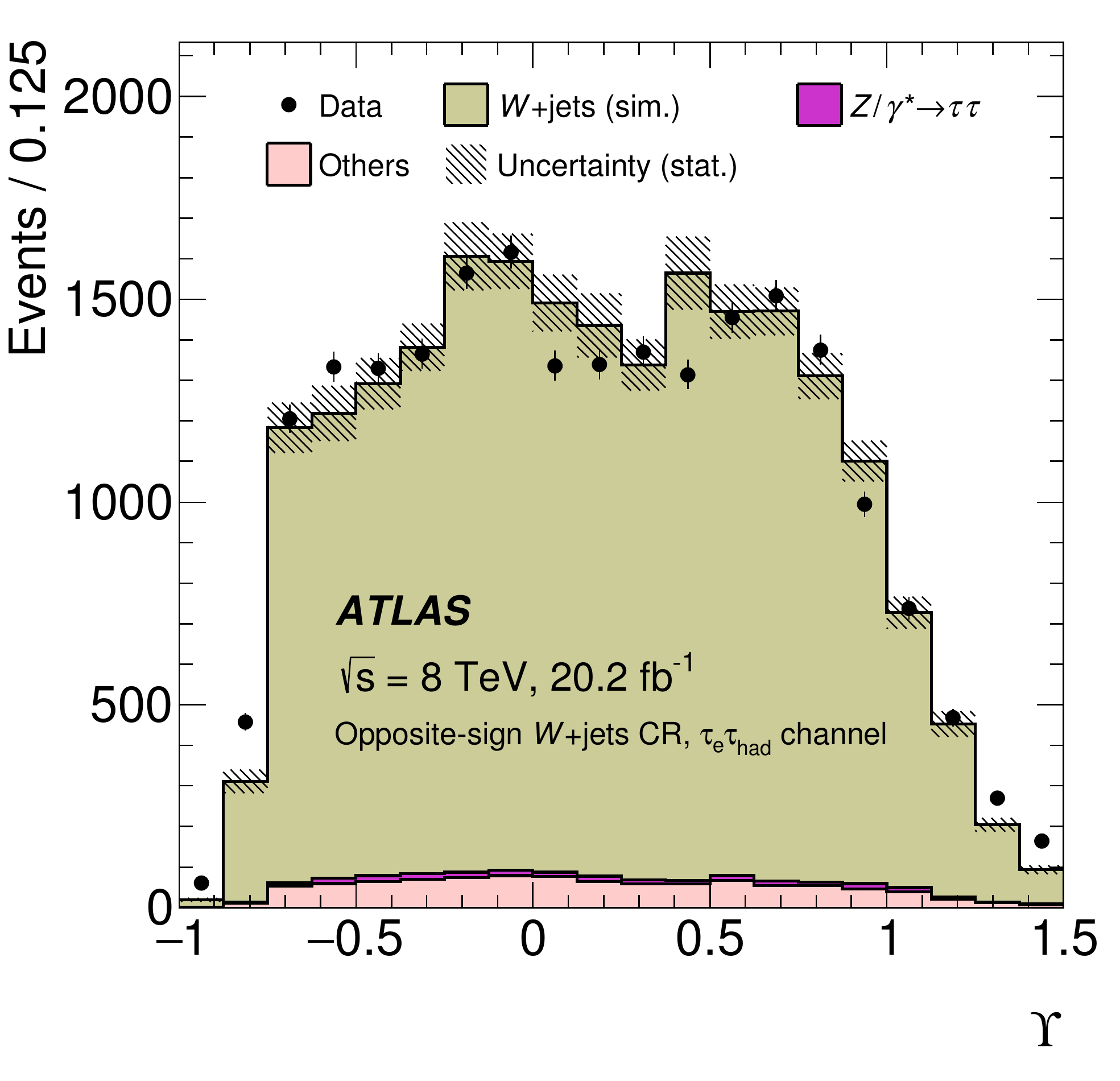}
   \includegraphics[width=0.48\textwidth,angle=0]{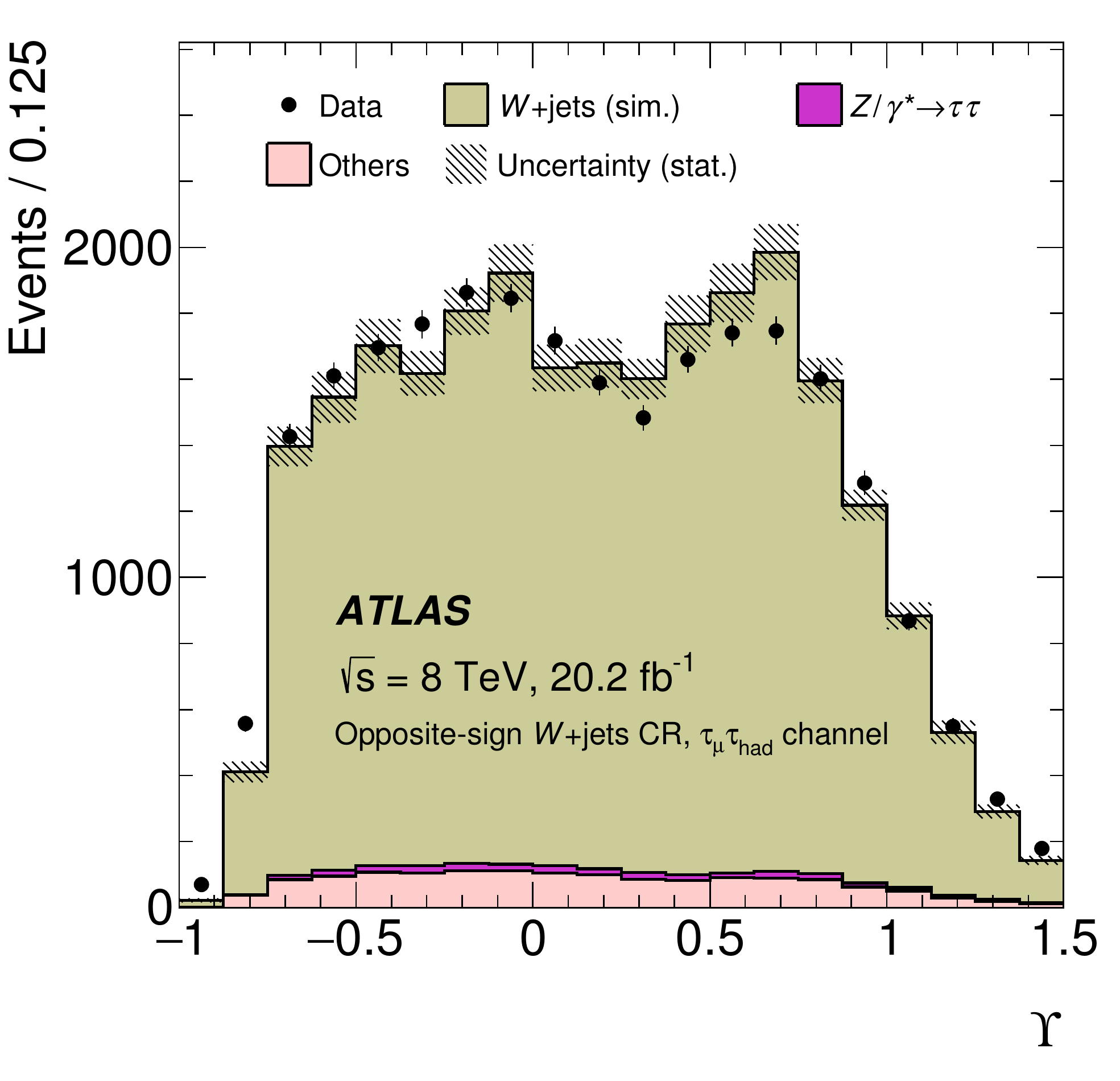}
}
\end{center}
\caption[]{
  The $\Upsilon$ distribution in the opposite-sign $W$+jets control region in the
  $\tau_e\text{--}\tauhad$ (left) and $\tau_\mu\text{--}\tauhad$ (right) channel. The contributions of
  $\Ztautau$ and of $\Zll$ and $\ttbar$ (other) events are estimated from the
  simulation. The $\tau$ polarisation in $\Ztautau$ events is obtained from the
  simulation. The shape of the $W$+jets contribution is estimated from the
  simulation as well. The $W$+jets contribution is normalised such that the
  total estimated event yield matches the observed yield. Only statistical
  uncertainties are shown. 
 \label{fig:WMCCR} }
\end{figure}

The $W$+jets background is estimated from a dedicated control region, which is
defined by inverting the $\sum\Delta\phi$ requirement applied in the signal
region selection and altering the transverse mass requirement to
$m_{\text{T}}>70$~GeV (see Table~\ref{tab:ControlRegionsOverview}).
Figure~\ref{fig:WMCCR} shows the $\Upsilon$ distribution in the $W$+jets
control region with data and simulation overlaid. 

Even though the simulation provides a reasonable description of the shape of the
$\Upsilon$ distribution in $W$+jets events, a more precise and robust
description is utilised. It is obtained from the large number of $W$+jets
events in the control region. For this, the $\Upsilon$ distribution for
$W$+jets events in the control region is estimated by subtracting the $\Zll$,
$\Ztautau$ and $\ttbar$ contributions as predicted by the simulation from the data. Here, the $\tau$
polarisation in $\Ztautau$ events is taken from the simulation. However, the
$W+$jets estimate is only negligibly affected if the $\tau$ polarisation in $\Ztautau$
events is assumed to be $-1$ or $+1$ instead of being taken from the simulation, because the
signal contamination in the $W$+jets control region is very small (below 1\%).  Due to the
strict transverse mass requirement, the multijet contribution in the $W$+jets
control region is negligible and it is thus ignored. 

Possible differences between the $\Upsilon$ distributions in $W$+jets events in
the $W$+jets control region and the selected signal region are assessed by performing a
linear fit to the ratio of these distributions in simulated $W$+jets events.
The fit functions describe the ratios within statistical uncertainties in both
channels. The resulting slopes are $0.03\pm0.05$ ($-0.02\pm0.05$) in the
$\tau_e\text{--}\tauhad$ ($\tau_\mu\text{--}\tauhad$) channel and are used to perform linear
corrections when transferring the $W+$jets $\Upsilon$ templates from the
$W+$jets control region to the selected signal region.

Additionally, the impact of altering the $\sum\Delta\phi$ and $m_{\mathrm{T}}$
requirements, which are used to define the $W+$jets control region, was studied
using dedicated validation regions. Differences between the $\Upsilon$ distributions
in the validation regions and the $W+$jets control region are evaluated using
additional linear fits. If one of the resulting slopes lies outside the range
covered by the statistical uncertainty in the slope estimated previously, the
uncertainty is inflated until the difference is covered. This results in an
inflation of the slope uncertainty in the $\tau_\mu\text{--}\tauhad$ channel by a
factor of $1.2$. The slope uncertainty in the $\tau_e\text{--}\tauhad$ channel remains
unchanged. The resulting uncertainties are referred to as $W$+jets shape uncertainties.

The normalisation of the $W$+jets contribution in the selected signal region is
determined by multiplying the event yield predicted from simulation by the
ratio of the $W$+jets event yields observed and predicted in the $W$+jets
control region.  The ratio is about $0.8$ in both channels.

An uncertainty of 3\% originates from the limited size of the simulated event samples
and is considered as a systematic uncertainty.

\subsection{Estimation of multijet background}
\label{sec:QCD}

The multijet background is estimated as follows. The shape of the $\Upsilon$
distribution is estimated from the same-sign control region, in which the
opposite-sign requirement on the lepton and $\tauhad$ candidates is reversed
(see Table~\ref{tab:ControlRegionsOverview}). The ratio \rQCD of multijet event
yields with opposite charge sign and same charge sign is used to scale the distribution obtained in the same-sign region. This ratio is measured in dedicated
multijet control regions in which the lepton isolation requirements are
inverted. 

In order to obtain the multijet contribution in the same-sign and multijet
control regions, the contributions from $W$+jets, $\Zll$, $\Ztautau$, and
$\ttbar$ events are subtracted from the data. These contributions
amount to about $28\%$ ($45\%$) of the data yield in
the same-sign region in the $\tau_e\text{--}\tauhad$ ($\tau_\mu\text{--}\tauhad$) channels and
to at most $16\%$ in the multijet control regions. The $\Zll$, $\Ztautau$,
and $\ttbar$ contributions are estimated and the $\tau$ polarisation in
$\Ztautau$ events is taken from the simulation. As in the $W+$jets
background estimate, an altered polarisation would have a negligible effect on
the multijet estimate. The $W+$jets contribution in the same-sign region is
estimated in the same way as in the selected signal region using the same-sign $W$+jets
control region.  The $W+$jets contribution in the opposite-sign (same-sign)
multijet control region is estimated as in the signal (same-sign) region. 

The value of \rQCD in the $\tau_e\text{--}\tauhad$ ($\tau_\mu\text{--}\tauhad$) channel is 1.05
(1.12), and the statistical uncertainty is negligible. The systematic
uncertainty is estimated by studying the dependence of the ratio of opposite-sign and same-sign event yields on the lepton isolation from well-isolated to
not isolated leptons. It is found to be $10\%$ ($9\%$) in the
$\tau_e\text{--}\tauhad$ ($\tau_\mu\text{--}\tauhad$) channel.

\begin{figure}[htbp]
  \begin{center}
{
   \includegraphics[width=0.48\textwidth,angle=0]{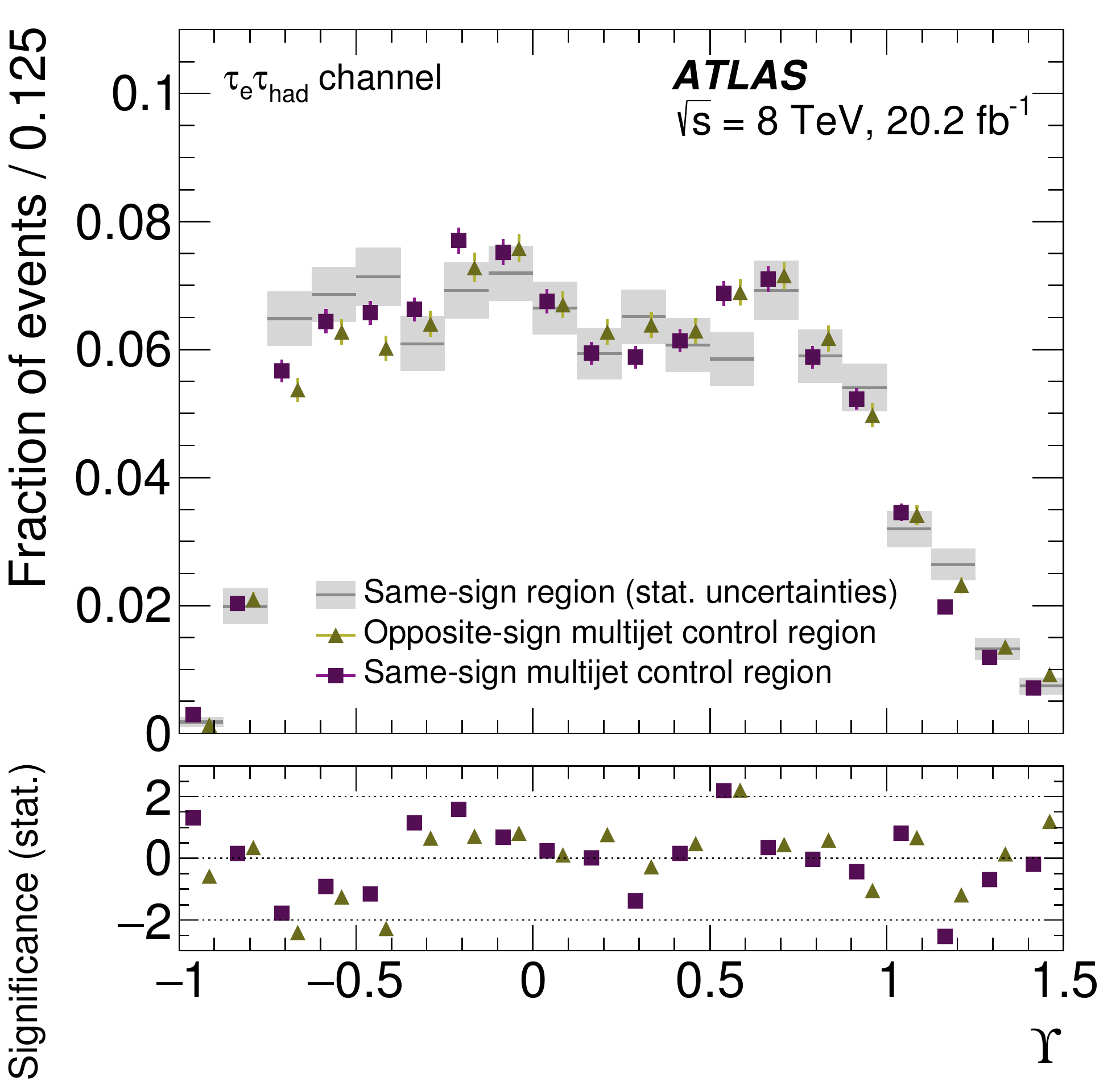}
   \includegraphics[width=0.48\textwidth,angle=0]{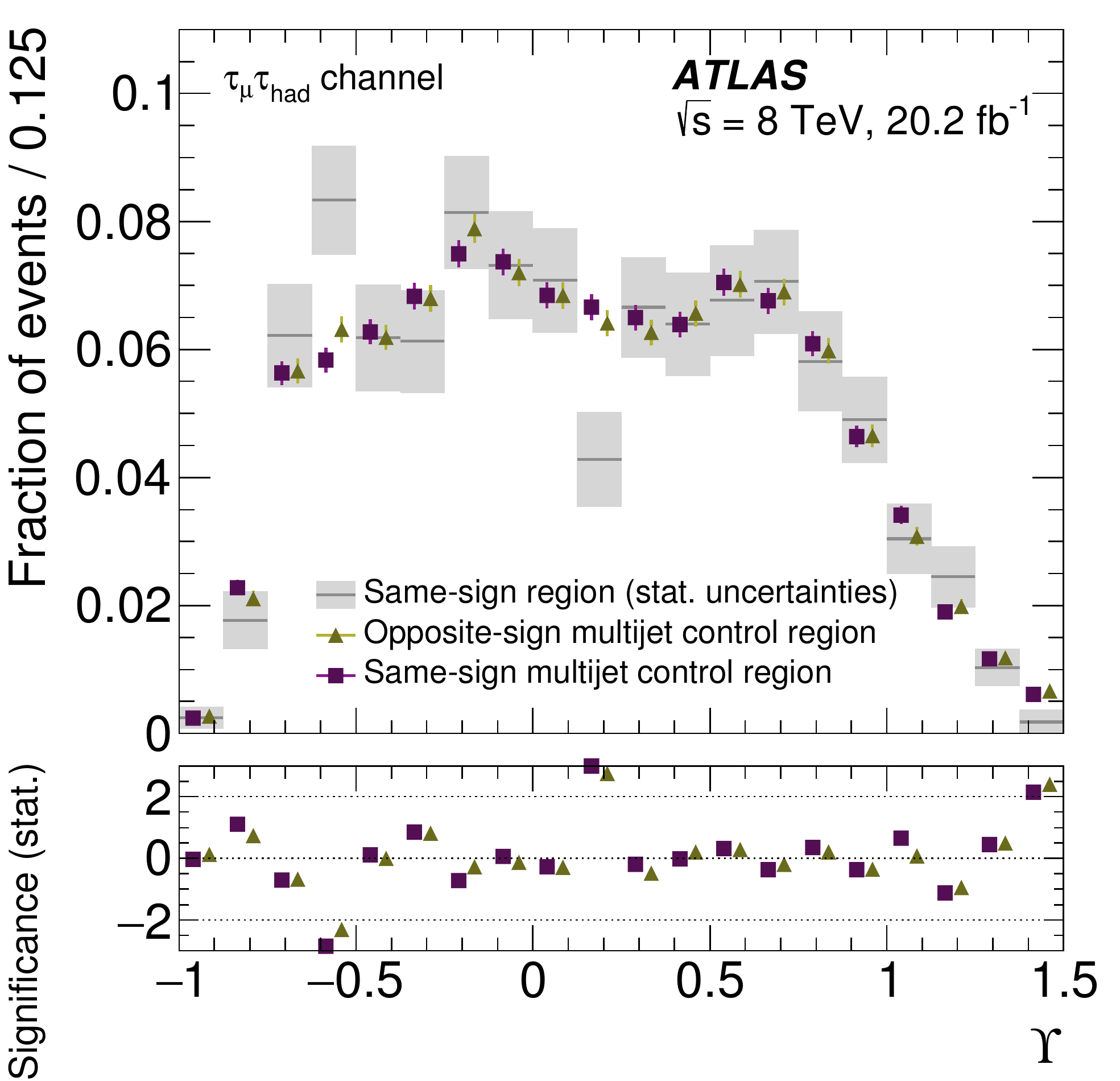}
}
\end{center}
\caption[]{
  Multijet templates obtained in the same-sign region and in the opposite- and
  same-sign multijet control regions in the $\tau_e\text{--}\tauhad$ (left) and
  $\tau_\mu\text{--}\tauhad$ (right) channel. Only statistical uncertainties are shown. 
  The significances, calculated from the statistical uncertainties, of the
  differences between the shapes in the same-sign region and those in the
  multijet control regions are shown as well.
 \label{fig:QCDshapes} }
\end{figure}

The multijet background estimate relies on the assumption that the shape of the $\Upsilon$
distribution is the same for multijet events with opposite and same sign lepton and $\tauhad$ candidates. This
is verified by comparing the distributions in the opposite-sign and same-sign
multijet control regions and in the same-sign region (see
Figure~\ref{fig:QCDshapes}). The shapes agree within the statistical
uncertainties in the same-sign region. 

\section{Systematic uncertainties}
\label{sec:syst}

The extraction of the $\tau$ polarisation from the observed data relies on the
prediction of the signal and background $\Upsilon$ templates. Systematic
uncertainties can affect the shape of the templates, as well as the acceptance
and thus the normalisation. The most important uncertainties are those that can
alter the shapes of the signal templates. 

\begin{figure}[htbp]
 \begin{center}
   \includegraphics[width=0.48\textwidth,angle=0]{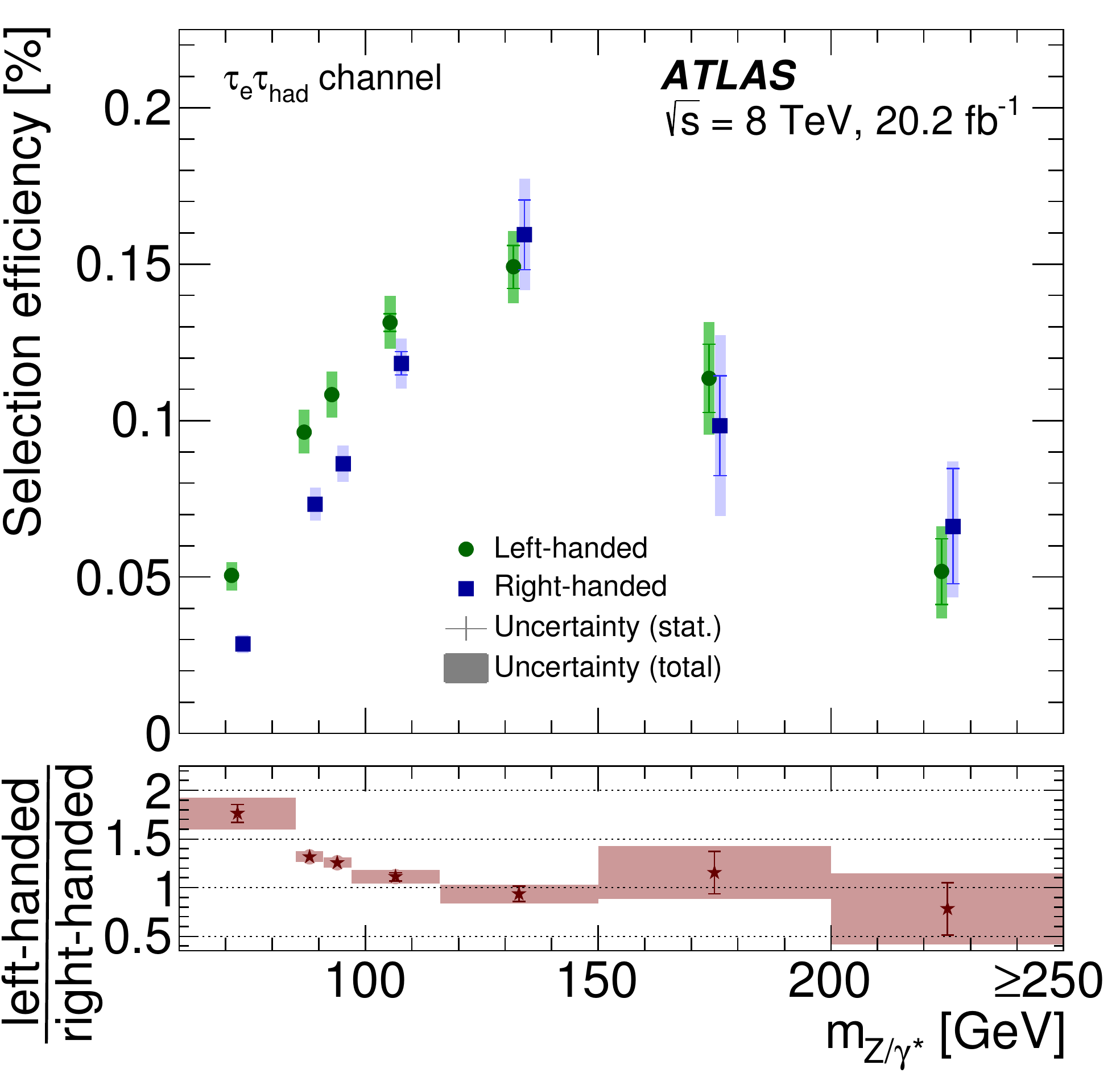}
   \includegraphics[width=0.48\textwidth,angle=0]{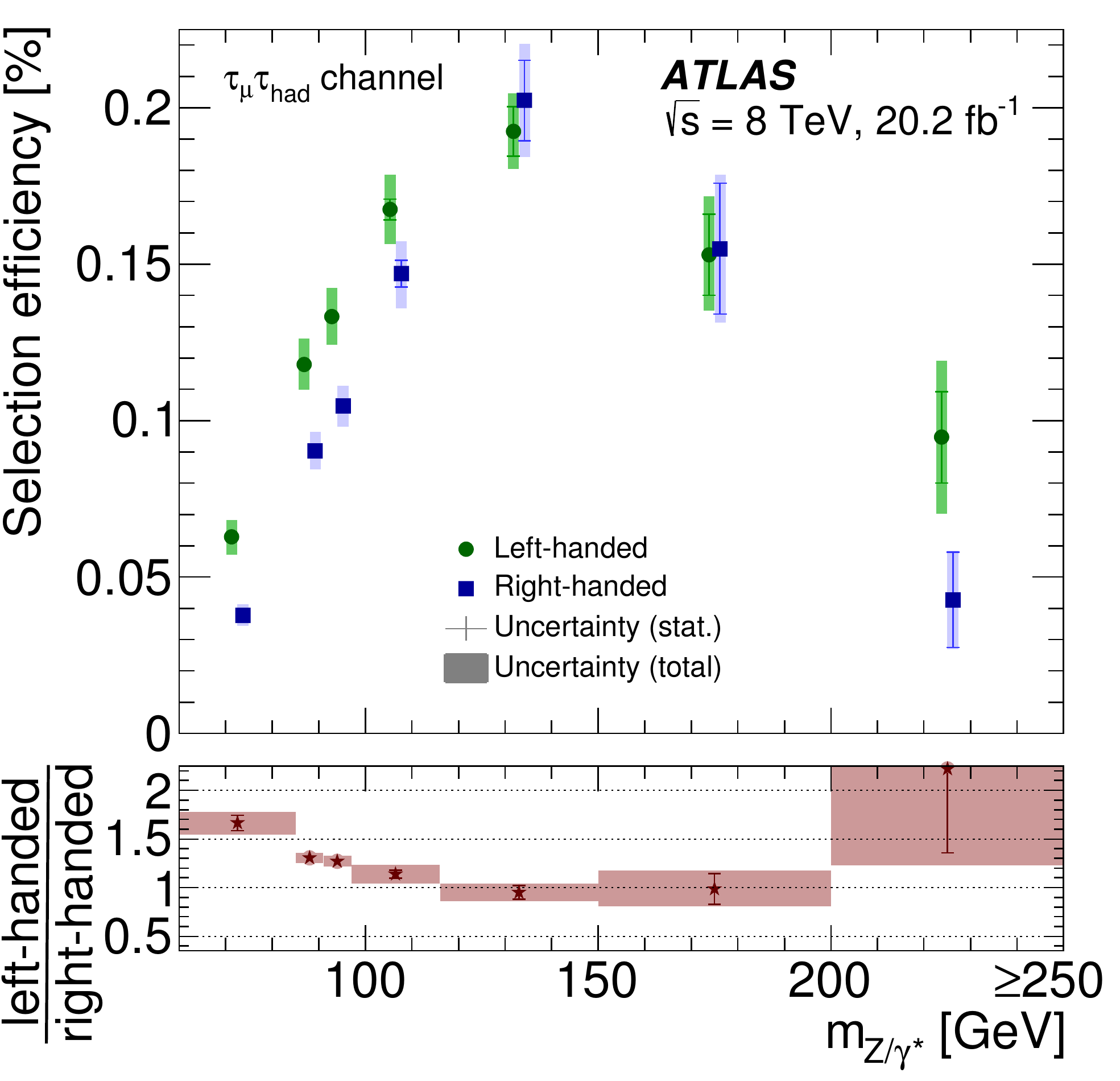}
 \end{center}

 \caption[]{
   Selection efficiency for signal events in the $\tau_e\text{--}\tauhad$ (left) and
   $\tau_\mu\text{--}\tauhad$ (right) channels as a function of $m_{\Zgamma}$. No
   requirement is placed on the $\tau$ decay modes at stable-particle level. The statistical
   and total uncertainties are indicated. The statistical and total
   uncertainties in the efficiency ratio are shown. The last bin includes overflow events.
 }
 \label{fig:selEff}
\end{figure}

Signal acceptance
uncertainties affect the left- and right-handed signal components in a very
similar way, which means that they have less impact in this analysis.  As the
background contamination is relatively small (about $30\%$ ($20\%$) in the
$\tau_e\text{--}\tauhad$ ($\tau_\mu\text{--}\tauhad$) channel), the systematic uncertainties
associated with its estimate have a minor impact on the measurement. The
uncertainties are discussed below, grouped into experimental and theoretical
uncertainties. Modelling uncertainties for the data-driven background estimates
are discussed in Section~\ref{sec:background}.  A detailed summary of the event
yields expected in the selected signal region with full uncertainties can be found in
Table~\ref{tab:SRyields}.  Figure~\ref{fig:selEff} shows the selection
efficiency of events with left- and right-handed $\tau^{-}$ as a function of
$m_{\Zgamma}$ for use in the interpretation of this measurement. Signal 
inefficiencies are dominated by decay mode and kinematic acceptance requirements.

\begin{table}[tb]
 \centering
  \sisetup{retain-explicit-plus, group-digits = false}
  \renewcommand{\arraystretch}{1.4}
   \begin{tabular}
  {l
    S[group-digits=false, table-format=9.0, table-number-alignment=right]@{$\,$}r
    S[group-digits=false, table-format=9.0, table-number-alignment=right,
    round-mode=figures, round-precision=2]@{$\,$}r}
    \toprule
    Process            & \multicolumn{2}{c}{$\tau_e\text{--}\tauhad$ channel}              & \multicolumn{2}{c}{$\tau_\mu\text{--}\tauhad$ channel}            \\
    \midrule
    Data               & \multicolumn{2}{c}{$32243$}           & \multicolumn{2}{c}{$32347$}                               \\
    Total expected     & 32000 & $\,^{\num{+1600}}_{\num{-1600}}$ & 32800 &$\,^{\num{+1800}}_{\num{-1800}}$ \\
    \midrule
    Left-handed        & 13800&$\,^{\num{+1100}}_{\num{-1100}}$ & 17000&$\,^{\num{+1400}}_{\num{-1300}}$ \\
    Right-handed       & 7800&$\,^{\num{+600}}_{\num{-600}}$    & 9600&$\,^{\num{+700}}_{\num{-700}}$    \\
    Outside mass-selected region & 430&$\,^{\num{+40}}_{\num{-40}}$       & 550&$\,^{\num{+40}}_{\num{-40}}$       \\
    $W+$jets           & 2240&$\,^{\num{+260}}_{\num{-240}}$    & 2590&$\,^{\num{+210}}_{\num{-220}}$    \\
    Multijet           & 6200&$\,^{\num{+600}}_{\num{-600}}$    & 2370&$\,^{\num{+270}}_{\num{-300}}$    \\
    Top pair             & 360&$\,^{\num{+40}}_{\num{-40}}$       & 390&$\,^{\num{+40}}_{\num{-40}}$       \\
    $(\Zll)+$jets        & 1210&$\,^{\num{+140}}_{\num{-140}}$    &  360&$\,^{\num{+50}}_{\num{-40}}$      \\
    \bottomrule
   \end{tabular}
 \caption{
  Event yields expected in the selected signal region for both channels. The $\Ztautau$
  contribution is shown separately for the three components used when
  extracting the polarisation in the 66--116~GeV mass-selected region (see
  Section~\ref{sec:selection}). The $\tau$ polarisation is assumed from the
  simulation for $\Ztautau$ events. Total uncertainties are shown.
 }
 \label{tab:SRyields}
\end{table}

\subsection{Experimental uncertainties}
\label{sec:syst_exp}

Experimental sources of uncertainty include trigger, object reconstruction and
identification efficiencies, energy and momentum scales and resolutions, and
the measurement of the integrated luminosity. They are described below in the
order of importance.

The efficiency for identifying $\tauhad$ candidates was measured in
data using tag-and-probe techniques and is about 55\% for single-prong $\tau$
leptons for the \enquote{medium} working point used in this
analysis~\cite{taureco}. The relative uncertainty in the $\tauhad$
identification efficiency is $(2\textnormal{--}3)\%$ for single-prong $\tau$
candidates. The simulated event samples are corrected for differences in the overall
efficiency between data and simulation and the associated uncertainties in the
normalisation of the signal and background templates are propagated through the
analysis. Some of the input variables~\cite{taureco} used in $\tauhad$
identification are strongly correlated with $\Upsilon$.  A mismodelling of
these input variables may thus cause differences between the shapes of the
$\Upsilon$ distributions in data and the simulation causing errors
specific to this analysis. These errors were studied in detail and are estimated by
comparing the $\tauhad$ identification input variable distributions of
$\tauhad$ candidates in $W$+jets and top pair events in the data and the
simulation. The observed differences are propagated through the analysis. The
resulting uncertainties are referred to as $\tauhad$ identification
uncertainties in the following.

The modelling of $\Upsilon$ strongly relies on the modelling of the energy
response to $\tauhad$, because the reconstructed $\tau$ energy is a direct
input (see Eq.~(\ref{eq:obs_exp})). In contrast to observables such as
masses of heavy particles, which are commonly exploited in analyses studying
decay channels that involve $\tau$ leptons, the reconstruction of $\Upsilon$ is
unaffected by the presence of neutrinos in $\tau$ decays. It is therefore
particularly sensitive to the modelling of the $\tauhadvis$ energy response.
Consequently, detailed studies were performed to provide a thorough
understanding of the related uncertainties. The Tau Energy Scale (TES) uncertainty for $\tauhad$
decays is evaluated based on the single-hadron response in the calorimeters
that was studied in Ref.~\cite{taureco}. The uncertainty is a
function of $\eta$ and $E_\textnormal{T}$ and is generally near $3\%$. A
mismodelling of the energy response to hadrons and to photons may affect the
$\Upsilon$ templates in different ways. For $\tauhad$ candidates with
$\Upsilon$ values around $+1$, most of the energy originates from hadrons, mostly
charged pions. Conversely, photons that arise from $\pi^0$ decays
typically carry a large fraction of the energy for candidates with $\Upsilon$
values close to $-1$. This is accounted for by splitting the TES uncertainty from
Ref.~\cite{taureco} into hadronic and electromagnetic components based on
the stable-particle-level fraction of the $\tauhadvis$ energy carried by hadrons and
photons, respectively, for signal events. 

A mismodelling of the $\tauhad$ energy resolution (TER) may affect the
modelling of the $\Upsilon$ distribution as well and may be distinguishable
from the effect caused by a mismodelling of the TES. The TER in ATLAS was not
measured before and is therefore evaluated in this analysis. The TER
uncertainties are considered for the hadronic and electromagnetic components
separately and determined from the $\Upsilon$ distribution in the same fit in
which the polarisation is measured. The absolute uncertainties are found to be
1.4\% for the hadronic and 1.8\% for the electromagnetic TER component.

The TES and TER uncertainties are each considered separately for the hadronic
and electromagnetic components. The TES uncertainty from the single-hadron
response studies is also considered for the backgrounds, which are estimated
from the simulation. Here, the contribution from $\Zee$ events, for which the
selected $\tauhad$ candidate originates from an electron, is treated separately
from the remaining backgrounds, for which the $\tauhad$ candidates originate
from quark- or gluon-initiated jets.

The remaining experimental uncertainties, referred to as other uncertainties, have a minor effect on the final
result:
\begin{itemize}

  \item Trigger, reconstruction and identification of electrons and muons: The
    efficiencies for triggering, reconstructing, and identifying electrons and
    muons are measured in data using tag-and-probe techniques.  Electron energy
    and muon momentum corrections and their uncertainties are evaluated
    by comparing the response in data and in the
    simulation~\cite{ATLAS-CONF-2014-032, PERF-2014-05}. The simulated event samples
    are corrected for the differences.

  \item Tag-and-probe studies of $\Zee$ events are used to derive the
    correction factors on the rate of electrons to be misidentified as
    $\tauhad$ leptons, as well their uncertainties~\cite{taureco}.

  \item Uncertainties that affect the $E_\textnormal{T}^\textnormal{miss}$
    estimation: In this analysis, uncertainties in the jet energy scale (JES)
    and resolution (JER) are only relevant due to their effect on the
    $E_\textnormal{T}^\textnormal{miss}$ reconstruction. Various sources of JES
    and JER uncertainty are considered~\cite{PERF-2012-01}. Along with the
    TES, TER, electron energy, and muon momentum uncertainties, they are
    propagated to the $E_\textnormal{T}^\textnormal{miss}$ calculation.  Additional uncertainties in the
    $E_\textnormal{T}^\textnormal{miss}$ scale and resolution due to energy
    clusters that do not belong to any reconstructed object are considered as well~\cite{PERF-2011-07}.

  \item Luminosity: The absolute luminosity scale is derived from
    beam-separation scans performed in November 2012. The uncertainty in the
    integrated luminosity is $1.9\%$~\cite{DAPR-2015-14}. It applies to simulated samples.

\end{itemize}
The uncertainties described above are propagated through the analysis.

\subsection{Theory uncertainties}\label{sec:theounc}

Theory uncertainties in the signal templates include uncertainties in the
event-by-event calculation of the helicity in the signal sample using the
\textsc{TauSpinner} algorithm, the choice of signal event generator and its
parton shower simulation model, and the choice of PDFs.

The uncertainty related to the signal sample splitting with the
\textsc{TauSpinner} algorithm is estimated by varying the relevant
\textsc{TauSpinner} input parameters. These are the QCD factorisation and
renormalisation scales, the $\alpha_\text{s}$ coupling and the PDFs.
Since the uncertainties may be mass dependent, they are calculated for three
different mass ranges around the $Z$ boson peak (66--116~GeV, 81--101~GeV, and 88--92~GeV). One of them coincides with the 66~$< m_{\Zgamma} <$~116~GeV range
used in this analysis. Samples of $pp \rightarrow \tau\tau+2$ jets events generated with the \textsc{MadGraph}~\cite{Alwall:2014hca} event generator interfaced with the
\textsc{Pythia8}~\cite{PYTHIA8} hadronisation and $\tau$ decay modelling and the same methods as in
Ref.~\cite{Richter-Was2016} are used. The signal samples used in the
analysis were generated with different $\sin^2\theta_\text{W}^\text{eff}$
values set in the \textsc{Alpgen} event generator and \textsc{Pythia6}+\textsc{Tauola} hadronisation and $\tau$ decay modelling. This may
result in an additional uncertainty in the sample splitting. To assess this
uncertainty, the polarisation obtained via the method described in
Section~\ref{sec:selection} is compared to the polarisation reported by the
\textsc{TauSpinner} algorithm. The difference is considered as a systematic
uncertainty. The two sources of signal sample splitting uncertainty have a similar
impact. Based on these studies, the signal template variations that are caused
by 1\% migrations from the left-handed to right-handed signal subsamples and
vice versa are considered and propagated through the analysis. The resulting 
uncertainties are referred to as signal sample splitting uncertainties.

The uncertainty related to the choice of event generator for the signal
sample is estimated with the help of two auxiliary samples produced with the
\textsc{Pythia8} event generator and with the \textsc{Powheg} event generator interfaced with the \textsc{Pythia8} hadronisation and $\tau$ decay modelling (see Table~\ref{tab:samples}). Because the
latter was generated using the CT10 PDF set, it is reweighted to match the
default one (CTEQ6L1) with the LHAPDF package \cite{lhapdf6} to avoid double-counting of possible systematic effects. These two
samples are used to obtain a set of event weights relative to the default
\textsc{Alpgen} sample before any event selection with respect to the
kinematics of $\tau$ leptons and $Z$ bosons and to the $\Upsilon$ spectra of various
hadronic $\tau$ decay modes.

The resulting uncertainties are among the leading ones in the analysis. Most of the impact
arises from the uncertainties in the $\tau$ lepton pseudorapidity distributions and from
the uncertainties in the $\Upsilon$ distributions in $\tau^{\pm}\rightarrow
h^{\pm}\pi^{0}\nu$ decays.
The estimation of the uncertainties related to the event generator in the
measurement of the polarisation in the fiducial region is performed in the same way as
described above. The uncertainties are referred to as signal modelling uncertainties.

The parton shower simulation model uncertainty is estimated using an auxiliary signal sample
produced with the \textsc{Alpgen} event generator interfaced the \textsc{Herwig} hadronisation modelling instead of the
\textsc{Pythia6} hadronisation modelling as in the default sample. 
It is used to obtain a set of event weights relative to the default \textsc{Alpgen} sample before
any event selection in the same way as for the uncertainties related to the event generator
choice described above. 

The impact of this systematic uncertainty, which is included in the other uncertainties category, on the final result is negligible.

The PDF-induced uncertainty is estimated by performing a reweighting of the signal
sample using the LHAPDF package. The nominal PDF set CTEQ6L1 is
reweighted to the following alternative LO PDF sets:
NNPDF30$\_$LO$\_$AS$\_$0118, MMHT2014LO68CL, and CT14LO. The uncertainties are
estimated for all three alternative PDF sets and found to be largest for the
CT14LO PDF set. The contribution of PDF uncertainties to the final polarisation
uncertainty is small.

\section{Fit model}
\label{sec:fit}

The $\tau$ polarisation is extracted in an extended binned maximum-likelihood
fit to the $\Upsilon$ distribution. The probability density function is
constructed in the histogram-based fitting tool HistFactory~\cite{HistFactory}
within the RooFit framework~\cite{RooFit}. The fit is performed simultaneously
in the signal and same-sign regions, each with 20 equally spaced bins in the
range $[-1,1.5]$ in $\Upsilon$, in both the $\tau_e\text{--}\tauhad$ and
$\tau_\mu\text{--}\tauhad$ channels. The fit to the observed data distribution is
performed twice, first to extract the $\tau$ polarisation in the range
of $66<m_{\Zgamma}<\SI{116}{\GeV}$ and then to measure the polarisation in
the fiducial region. 

The signal histograms of the $\Upsilon$ variable for the fit that extracts the
polarisation in the mass-selected region are the respective three $\Ztautau$
contributions (see Section~\ref{sec:selection}) that pass the selected signal region and
same-sign region event selections in the simulation. They are passed to the fit
as nominal signal templates. The left-handed and right-handed signal templates
describing events inside the mass-selected region are each normalised to the full
$\Ztautau$ cross-section inside the mass-selected region. The relative contributions are
scaled with the parameter of interest, $\Ptaupar$, such that $\Ptaupar$
represents the polarisation at production as defined in Eq.~(\ref{eqn:ptau})
without any selection except the $66<m_{\Zgamma}<\SI{116}{\GeV}$
requirement. The template for $\Ztautau$ events outside the mass-selected region is
scaled with the respective $\Ztautau$ cross-section and is not affected by the
parameter $\Ptaupar$. Effects causing deviations of the expected polarisation
from that in the data could also alter the $\Ztautau$ normalisation. Hence an
additional unconstrained fit parameter, $\alpha_Z$, is included to scale the
overall normalisation of the $\Ztautau$ signals. The $\Ptaupar$ and $\alpha_Z$
parameters are common to the fitted relative and overall normalisation of the
signal templates in all regions.

The signal templates used in the fit that extracts the $\tau$ polarisation in the
fiducial region are obtained in a similar way using the respective three
contributions defined in Section~\ref{sec:selection}. Here, the left- and
right-handed signal templates corresponding to events inside the fiducial
region are each scaled with the full $\Ztautau$ cross-section inside the
fiducial region. Due to this scaling $\Ptaupar$ then represents the
polarisation of $\tau$ leptons produced in the fiducial region. The
contribution made by events outside the fiducial region is treated as
previously described for the events outside the mass-selected region. The scaling with
$\Ptaupar$ and $\alpha_Z$ is also done as described for the mass-selected region. The treatment of
the backgrounds and systematic uncertainties is described below.

The $Z$+jets and $\ttbar$ backgrounds are taken into account by the simulated
$\Upsilon$ distributions passing the selected signal region and same-sign region event
selections. The $W$+jets template histograms are taken from the data-driven
estimate. Each of the $Z$+jets, $W$+jets, and $\ttbar$ background templates
are normalised to the expected number of events for each background in the
respective regions as described in Section~\ref{sec:background}. The multijet
background is estimated in a simultaneous fit in the signal and same-sign
regions with nuisance parameters common to the two regions per bin and channel
to fit the content in each. The related uncertainties are referred to as multijet estimate uncertainties.
For each channel the normalisation of the multijet
background in the selected signal region relative to the same-sign region is scaled via
a fixed normalisation parameter, \rQCD.

A summary of the nuisance parameters related to systematic uncertainties can be
found in Table~\ref{tab:summary_NPs}. All systematic uncertainties in the same-sign region are much smaller than the statistical uncertainties in the multijet
estimate. They are thus negligible and omitted in the fit.

\begin{table}[tb]
\begin{center}
\begin{tabular}{l
                S[group-digits=false, table-format=9.0, table-number-alignment=right,
                round-mode=figures, round-precision=2]@{$\,$}r
                l
                l
               }

  \toprule
  Source of uncertainty       & \multicolumn{2}{c}{Number of parameters}  & Constraint  & Steer variation of     \\
  \midrule
  Multijet estimate           & 40                  &  & None       & one bin each             \\
  \midrule
  MC statistical              & 40                  &  & Poissonian & one bin each             \\
  \midrule
  Modelling of signal process & 3                   &  & Gaussian   & shape and normalisation  \\
  $\tauhad$ identification    & 5                   &  & Gaussian   & shape or normalisation   \\
  Signal sample splitting     & 2                   &  & Gaussian   & shape and normalisation  \\
  TES and TER                 & 6                   &  & Gaussian   & shape and normalisation  \\
  PDF                         & 1                   &  & Gaussian   & shape and normalisation  \\
  $W$+jets shape              & 2                   &  & Gaussian   & shape                    \\
  Other                       & 34~or~36            &  & Gaussian   & normalisation            \\
  \bottomrule
\end{tabular}
\end{center}

\caption{
  Summary of nuisance parameters related to systematic uncertainties considered
  in the fits that extract the $\tau$ polarisation when combining the two
  channels. The number of parameters in the \enquote{Other} category is $36$
  ($34$) in the fit that extracts the polarisation in the mass-selected
  region (in the fiducial region). \label{tab:summary_NPs}
}
\end{table}

The statistical uncertainty associated with the finite size of the simulated event
samples is accounted for with a variation of the Barlow--Beeston treatment~\cite{barlowbeeston}.
This results in one nuisance parameter per channel and bin. The related uncertainties are referred to as MC statistical uncertainties.

Further nuisance parameters are included to account for systematic variations
of the template shape and normalisation estimated with the methods described in
Sections~\ref{sec:background} and~\ref{sec:syst}. The systematic uncertainties
are accounted for in the fit with variations of the individual nominal template
histograms. These variations may change the overall normalisation of the
histogram or may introduce bin-dependent shape differences. In either case, a
single nuisance parameter interpolates between variations that correspond to
the estimated $+1\sigma$ and $-1\sigma$ uncertainties with a Gaussian
constraint. The nuisance parameters may be correlated between normalisation and
shape variations, between samples, regions, and channels.

The signal process modelling and PDF parameters control the variations
introduced when changing the event generator or the PDF set, respectively.
Three of the parameters related to $\tauhad$ identification uncertainties account for
the systematic variation of the input variables that may significantly affect
the modelling of the signal template shapes in the simulation. The remaining
two $\tauhad$ identification parameters exclusively vary the normalisation of the
signal and background templates according to the uncertainties estimated in the
tag-and-probe studies from Ref.~\cite{taureco}. The correlations of the
normalisation and shape uncertainties are not known and the parameters are
treated as uncorrelated. It was verified that the correlation assumed has a
negligible effect on the overall uncertainty. One parameter controls each of
the variations caused by migrations from the left-handed to right-handed
signal subsamples and vice versa, accounting for the signal sample splitting
uncertainties. The correlations of these parameters are also unknown. They are
treated as uncorrelated. Their impact on the polarisation uncertainty
would be reduced, if they were assumed to be fully correlated instead. One parameter
controls each of the variations of the hadronic and electromagnetic components
of the TES and TER in the signal templates. The remaining TES parameters
account for the TES uncertainty in the backgrounds. One of them is dedicated to
$\Zee$ events, in which one of the electrons is misidentified as a $\tauhad$ candidate. One $W$+jets shape parameter per channel accounts for the shape
uncertainties described in Section~\ref{sec:Wjets}.

The remaining systematic uncertainties are considered for their impact on the
normalisation of each of the template histograms. 

Most of them have a small impact on the templates, individually. In the signal
region for each sample, the systematic uncertainties are ordered by decreasing
amount of normalisation variation that they cause. Nuisance parameters are
included until at least $95\%$ of the sum of all normalisation uncertainties
per sample is covered. It was verified that the remaining uncertainties would
have a negligible impact, if considered.

The fit model was validated in detail using pseudo-experiments. It was verified that it
correctly determines the polarisation when confronted with data samples that include polarisation values
different from those found in the simulation. The bias was found negligible and the uncertainties determined
by the fit were found accurate.

\section{Results}
\label{sec:result}

\begin{figure}[htbp]
 \begin{center}
   \includegraphics[width=0.48\textwidth,angle=0]{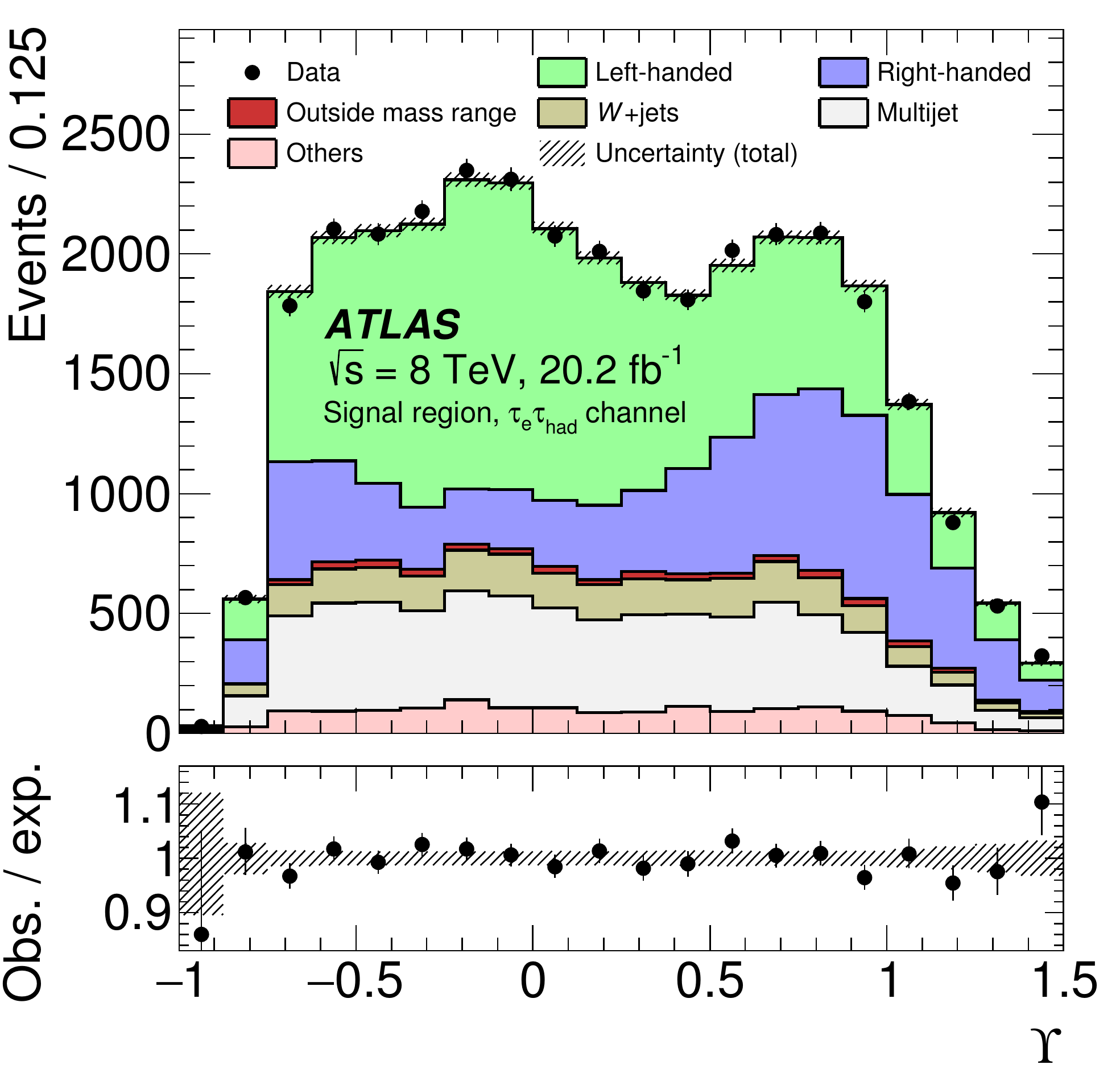}
   \includegraphics[width=0.48\textwidth,angle=0]{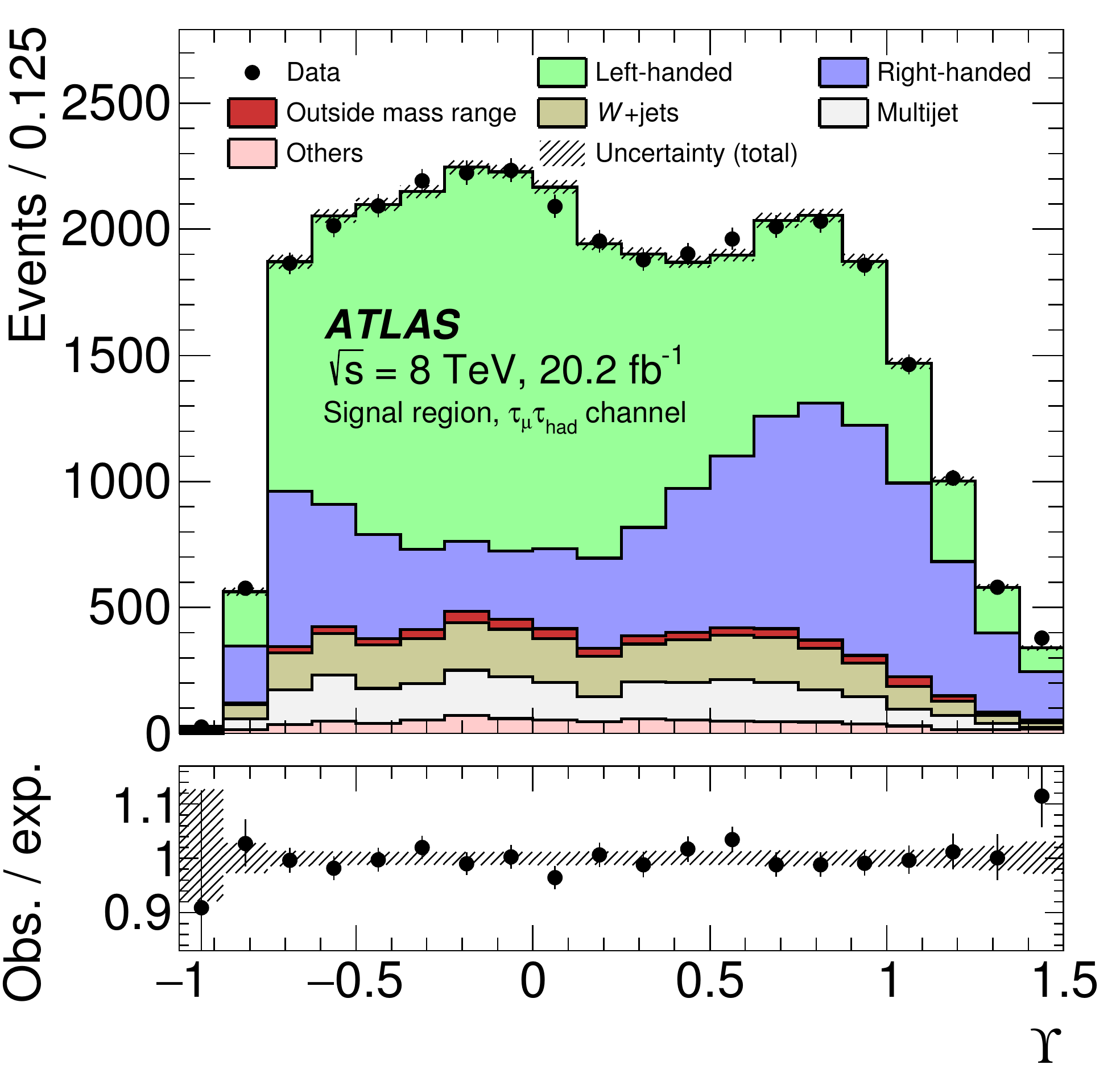}
   \includegraphics[width=0.48\textwidth,angle=0]{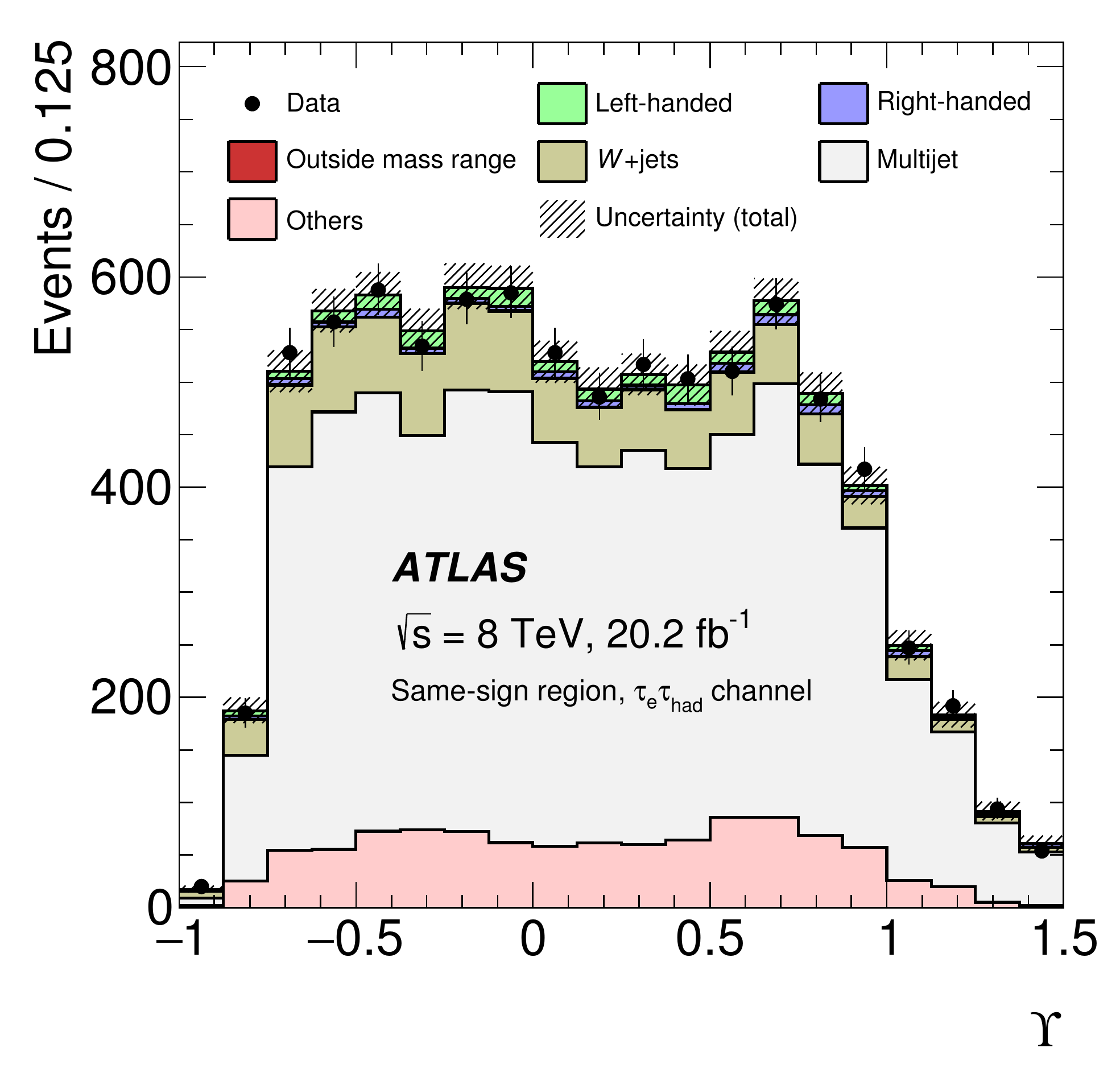}
   \includegraphics[width=0.48\textwidth,angle=0]{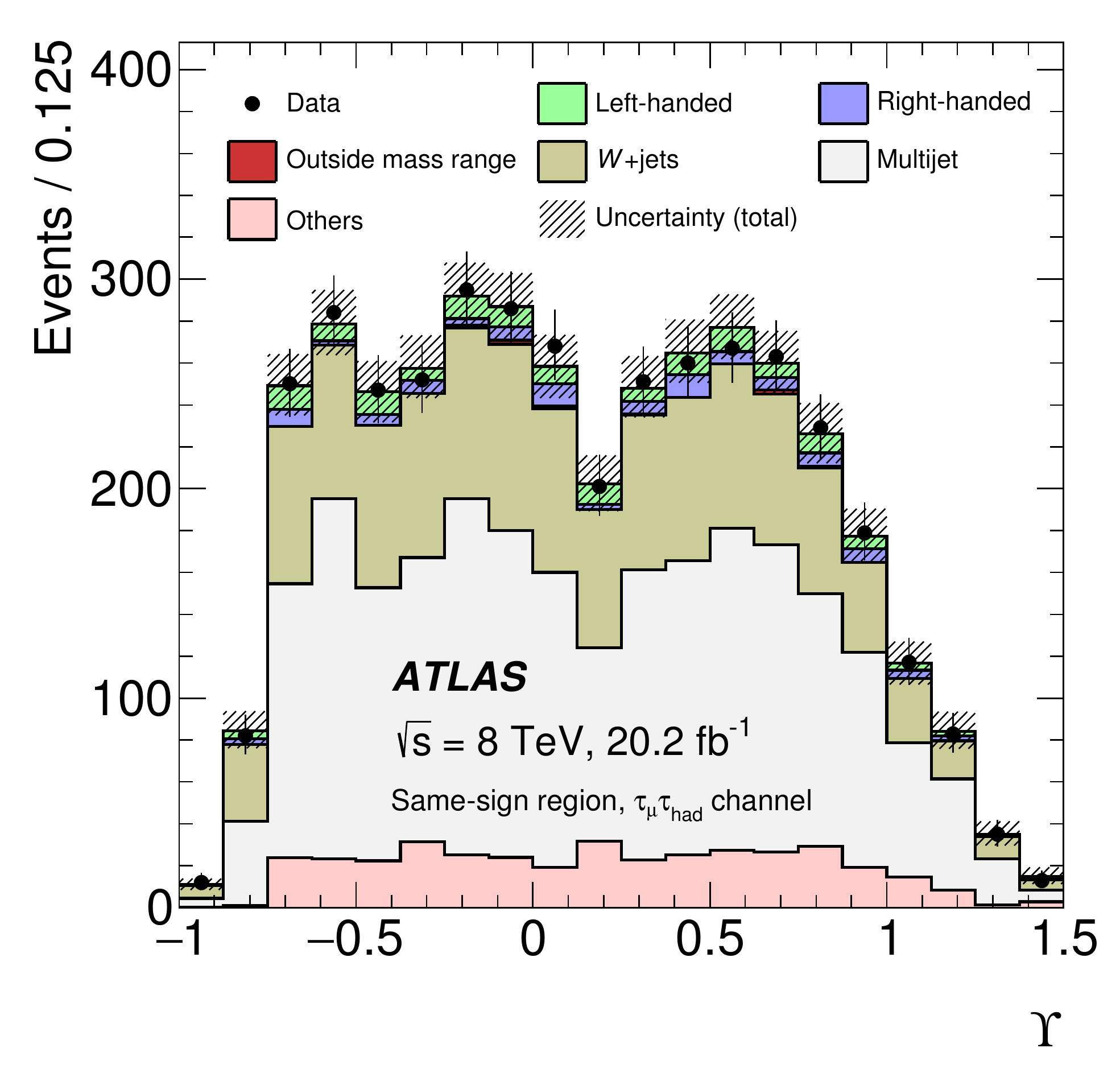}
 \end{center}

 \caption[]{
  Post-fit $\Upsilon$ distributions for the $\tau_e\text{--}\tauhad$ (left) and
  $\tau_\mu\text{--}\tauhad$ (right) channels, and for the signal (top) and same-sign
  (bottom) regions for the fit that extracts the $\tau$ polarisation in the
  mass-selected region of $66<m_{\Zgamma}<\SI{116}{\GeV}$.
 }
 \label{fig:postCombinedFit}
\end{figure}

\begin{figure}[htbp]
 \begin{center}
   \includegraphics[width=0.48\textwidth,angle=0]{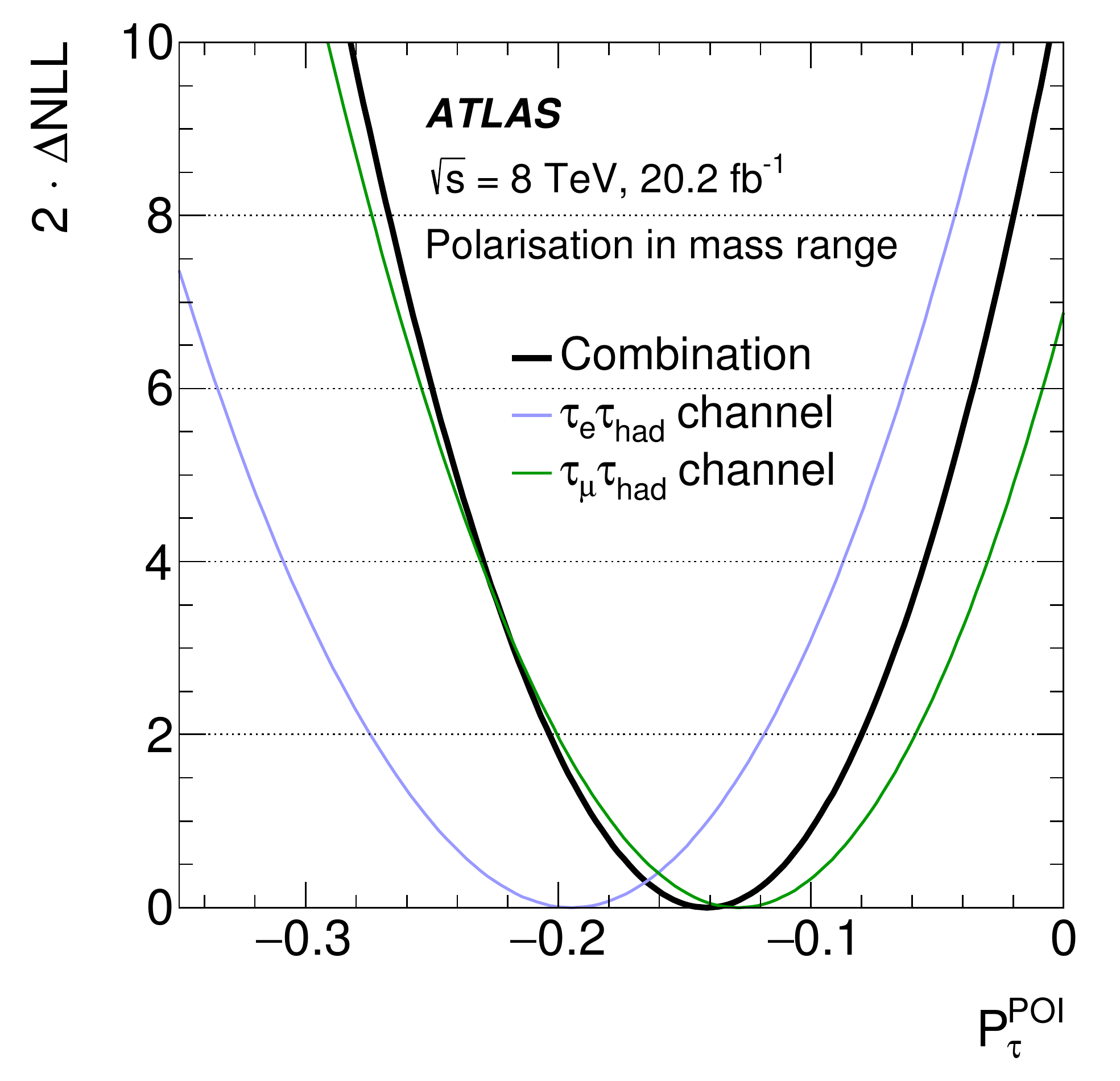}
   \includegraphics[width=0.48\textwidth,angle=0]{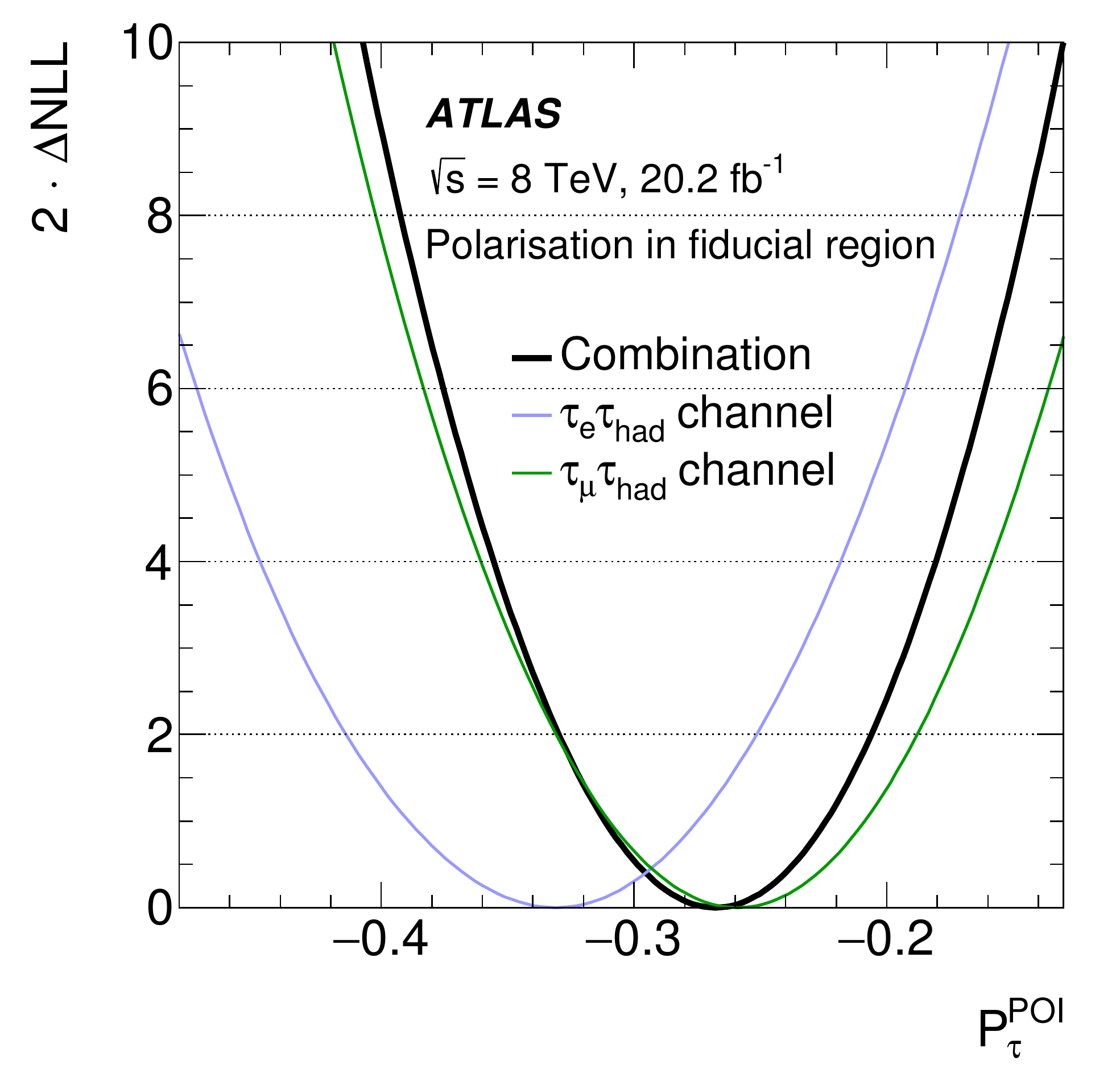}
 \end{center}
 
 \caption[]{
    Likelihood profiles of $\Ptaupar$ for the fits that extract the polarisation
in the mass-selected region of $66<m_{\Zgamma}<\SI{116}{\GeV}$ (left) and in
the fiducial region (right). The profiles are shown separately for the fits in
the $\tau_e\text{--}\tauhad$ and $\tau_\mu\text{--}\tauhad$ channels and for the combination.
 }
 \label{fig:LLHprofiles}
\end{figure}

The $\tau$ polarisations in the mass-selected region of
$66<m_{\Zgamma}<\SI{116}{\GeV}$, and in the fiducial region, are
extracted using the extended, binned maximum-likelihood fit described in
Section~\ref{sec:fit}. The fit is performed for the individual channels and for
the combination. The $\Upsilon$ distributions after the combined fit that
extracts the $\tau$ polarisation in the mass-selected region are shown in
Figure~\ref{fig:postCombinedFit}. The $\Ptaupar$ likelihood profiles are shown
in Figure~\ref{fig:LLHprofiles} and the resulting polarisation values are
summarised in Table~\ref{tab:fitResults}. The polarisation values measured in
the $\tau_e\text{--}\tauhad$ and $\tau_\mu\text{--}\tauhad$ channels agree at a level of 1.4 standard
deviations and are compatible. Only uncertainties that are uncorrelated between the channels
are considered in this compatibility estimate. Apart from the statistical
uncertainties, these are the uncertainties related to the finite size of the
simulated event samples and those related to the multijet background estimate. Some
of the nuisance parameters, which correspond to uncertainties that are specific
to this analysis such as the uncertainties in the modelling of $\tauhad$ identification and 
$\tauhad$ energy reconstruction on the 
$\Upsilon$ distribution, are fit to values that differ from their nominal estimates.
The sizes of these \enquote{pulls} are similar in the two channels. The largest effect 
is that the polarisation value obtained in the combination is higher and close to 
that measured in the $\tau_\mu\text{--}\tauhad$ channel.

The impact of the different sources of
uncertainty is summarised in Table~\ref{tab:uncertatiesImpact}.

The uncertainty in a $\sin^2\theta_\text{W}^\text{eff}$ value extracted from
this measurement would be approximately 15 times larger than that reached by
the LEP experiments from $\tau$ polarisation~\cite{Ptau_LEP_new}. Therefore, and
because additional studies would be required to correct for the $Z$ boson and photon
interference, $\sin^2\theta_\text{W}^\text{eff}$ is not determined here.

\begin{table}[tb]
  \begin{center}
  \sisetup{retain-explicit-plus}
 \begin{tabular}{
lr@{\,}@{$\pm$}@{\,}l@{\,}l
r@{\,}@{$\pm$}@{\,}l@{\,}l
}
  \toprule
  Channel                       & \multicolumn{3}{c}{$\Ptau$ in mass-selected region}      & \multicolumn{3}{c}{$\Ptau$ in fiducial region} \\
  \midrule
  $\tau_e\text{--}\tauhad$      & $-0.20$ &$0.02\;(\text{stat})$&$\pm\;0.05\;(\text{syst})$ & $-0.33$&$0.03\;(\text{stat})$&$\pm\;0.05\;(\text{syst})$ \\
  $\tau_\mu\text{--}\tauhad$    & $-0.13$ &$0.02\;(\text{stat})$&$\pm\;0.05\;(\text{syst})$ & $-0.26$&$0.02\;(\text{stat})$&$\pm\;0.05\;(\text{syst})$ \\
  \midrule
  Combination                   & $-0.14$ &$0.02\;(\text{stat})$&$\pm\;0.04\;(\text{syst})$ & $-0.27$&$0.02\;(\text{stat})$&$\pm\;0.04\;(\text{syst})$ \\
  \bottomrule
 \end{tabular}
  \end{center}
 \caption{
  Measured $\tau$ polarisation values and overall uncertainties in the mass-selected region of
  $66<m_{\Zgamma}<\SI{116}{\GeV}$ and in the fiducial region.
 }\label{tab:fitResults}
\end{table}

\begin{table}[tb]
  \begin{center}
  \sisetup{retain-explicit-plus}
 \begin{tabular}{lcc}
  \toprule
  Source of uncertainty         & $\sigma_{\Ptau}$ in mass-selected region   & $\sigma_{\Ptau}$ in fiducial region \\
  \midrule
  Modelling of signal process   & $0.026$ & $0.022$ \\
  $\tauhad$ identification      & $0.020$ & $0.024$ \\
  MC statistical                & $0.016$ & $0.019$ \\
  Signal sample splitting       & $0.015$ & $0.015$ \\
  TES and TER                   & $0.015$ & $0.019$ \\
  Multijet estimate             & $0.013$ & $0.013$ \\
  PDF                           & $0.007$ & $0.005$ \\
  $W$+jets shape                & $0.002$ & $0.003$ \\
  Other                         & $0.008$ & $0.003$ \\
  \midrule
  Total systematic uncertainty  & $0.040$ & $0.039$ \\
  Statistical uncertainty       & $0.015$ & $0.016$ \\
  \bottomrule
 \end{tabular}
  \end{center}
 \caption{
    Impact of the individual sources of uncertainty on the polarisation
    uncertainty $\sigma_{\Ptau}$ in the combined fits that extract the $\tau$
    polarisation in the mass-selected region of $66<m_{\Zgamma}<\SI{116}{\GeV}$
    and in the fiducial region.  The total systematic uncertainty quoted is
    estimated from the total uncertainty and the statistical uncertainty.
    }
    \label{tab:uncertatiesImpact}
 
\end{table}

\FloatBarrier

\section{Conclusion}
\label{sec:conclusion}

A measurement of the $\tau$ polarisation in $\Ztautau$ decays with one leptonic
and one single-prong hadronic $\tau$ decay is performed. Sensitivity to $\tau$
polarisation is gained from the hadronic $\tau$ decay.  The 20.2~$\ifb$ dataset
of proton--proton collisions at $\sqrt{s}=8$~TeV collected by the ATLAS
experiment at the LHC in 2012 is utilised. The measurement is complementary to
previous measurements in electron--positron collisions.

In the fiducial region, the measured $\tau$ polarisation is
$\Ptau~=-0.27\;\pm~0.02\;(\text{stat})\;\pm~0.04\;(\text{syst})$.
It agrees with the value predicted by the Standard Model (as implemented in the
\textsc{Alpgen} event generator interfaced with the \textsc{Pythia6} and
\textsc{Tauola}) hadronisation and $\tau$ decay modelling, which is $\Ptau~=-0.270\;\pm\;0.006$.  The polarisation is
then extracted in the mass-selected region of 66~$< m_{\Zgamma} <$~116~GeV and a
value of $\Ptau~=-0.14\;\pm~0.02\;(\text{stat})\;\pm~0.04\; (\text{syst})$ is
found. The result is in agreement with Standard Model prediction of
$\Ptau~=-0.1517\;\pm\;0.0019$. 

\section*{Acknowledgements}

We thank CERN for the very successful operation of the LHC, as well as the
support staff from our institutions without whom ATLAS could not be
operated efficiently.

We acknowledge the support of ANPCyT, Argentina; YerPhI, Armenia; ARC, Australia; BMWFW and FWF, Austria; ANAS, Azerbaijan; SSTC, Belarus; CNPq and FAPESP, Brazil; NSERC, NRC and CFI, Canada; CERN; CONICYT, Chile; CAS, MOST and NSFC, China; COLCIENCIAS, Colombia; MSMT CR, MPO CR and VSC CR, Czech Republic; DNRF and DNSRC, Denmark; IN2P3-CNRS, CEA-DRF/IRFU, France; SRNSF, Georgia; BMBF, HGF, and MPG, Germany; GSRT, Greece; RGC, Hong Kong SAR, China; ISF, I-CORE and Benoziyo Center, Israel; INFN, Italy; MEXT and JSPS, Japan; CNRST, Morocco; NWO, Netherlands; RCN, Norway; MNiSW and NCN, Poland; FCT, Portugal; MNE/IFA, Romania; MES of Russia and NRC KI, Russian Federation; JINR; MESTD, Serbia; MSSR, Slovakia; ARRS and MIZ\v{S}, Slovenia; DST/NRF, South Africa; MINECO, Spain; SRC and Wallenberg Foundation, Sweden; SERI, SNSF and Cantons of Bern and Geneva, Switzerland; MOST, Taiwan; TAEK, Turkey; STFC, United Kingdom; DOE and NSF, United States of America. In addition, individual groups and members have received support from BCKDF, the Canada Council, CANARIE, CRC, Compute Canada, FQRNT, and the Ontario Innovation Trust, Canada; EPLANET, ERC, ERDF, FP7, Horizon 2020 and Marie Sk{\l}odowska-Curie Actions, European Union; Investissements d'Avenir Labex and Idex, ANR, R{\'e}gion Auvergne and Fondation Partager le Savoir, France; DFG and AvH Foundation, Germany; Herakleitos, Thales and Aristeia programmes co-financed by EU-ESF and the Greek NSRF; BSF, GIF and Minerva, Israel; BRF, Norway; CERCA Programme Generalitat de Catalunya, Generalitat Valenciana, Spain; the Royal Society and Leverhulme Trust, United Kingdom.

The crucial computing support from all WLCG partners is acknowledged gratefully, in particular from CERN, the ATLAS Tier-1 facilities at TRIUMF (Canada), NDGF (Denmark, Norway, Sweden), CC-IN2P3 (France), KIT/GridKA (Germany), INFN-CNAF (Italy), NL-T1 (Netherlands), PIC (Spain), ASGC (Taiwan), RAL (UK) and BNL (USA), the Tier-2 facilities worldwide and large non-WLCG resource providers. Major contributors of computing resources are listed in Ref.~\cite{ATL-GEN-PUB-2016-002}.

\printbibliography

\newpage 
\begin{flushleft}
{\Large The ATLAS Collaboration}

\bigskip

M.~Aaboud$^\textrm{\scriptsize 137d}$,
G.~Aad$^\textrm{\scriptsize 88}$,
B.~Abbott$^\textrm{\scriptsize 115}$,
O.~Abdinov$^\textrm{\scriptsize 12}$$^{,*}$,
B.~Abeloos$^\textrm{\scriptsize 119}$,
S.H.~Abidi$^\textrm{\scriptsize 161}$,
O.S.~AbouZeid$^\textrm{\scriptsize 139}$,
N.L.~Abraham$^\textrm{\scriptsize 151}$,
H.~Abramowicz$^\textrm{\scriptsize 155}$,
H.~Abreu$^\textrm{\scriptsize 154}$
% ...
\bigskip
\\
% ...
$^{12}$ Institute of Physics, Azerbaijan Academy of Sciences, Baku, Azerbaijan\\
% ...
$^{88}$ CPPM, Aix-Marseille Universit{\'e} and CNRS/IN2P3, Marseille, France\\
% ...
$^{115}$ Homer L. Dodge Department of Physics and Astronomy, University of Oklahoma, Norman OK, United States of America\\
% ...
$^{119}$ LAL, Univ. Paris-Sud, CNRS/IN2P3, Universit{\'e} Paris-Saclay, Orsay, France\\
% ...
$^{137}$ $^{(a)}$ Facult{\'e} des Sciences Ain Chock, R{\'e}seau Universitaire de Physique des Hautes Energies - Universit{\'e} Hassan II, Casablanca; $^{(b)}$ Centre National de l'Energie des Sciences Techniques Nucleaires, Rabat; $^{(c)}$ Facult{\'e} des Sciences Semlalia, Universit{\'e} Cadi Ayyad, LPHEA-Marrakech; $^{(d)}$ Facult{\'e} des Sciences, Universit{\'e} Mohamed Premier and LPTPM, Oujda; $^{(e)}$ Facult{\'e} des sciences, Universit{\'e} Mohammed V, Rabat, Morocco\\
% ...
$^{139}$ Santa Cruz Institute for Particle Physics, University of California Santa Cruz, Santa Cruz CA, United States of America\\
% ...
$^{151}$ Department of Physics and Astronomy, University of Sussex, Brighton, United Kingdom\\
% ...
$^{154}$ Department of Physics, Technion: Israel Institute of Technology, Haifa, Israel\\
$^{155}$ Raymond and Beverly Sackler School of Physics and Astronomy, Tel Aviv University, Tel Aviv, Israel\\
% ...
$^{161}$ Department of Physics, University of Toronto, Toronto ON, Canada\\
% ...
$^{a}$ Also at Department of Physics, King's College London, London, United Kingdom\\
$^{b}$ Also at Institute of Physics, Azerbaijan Academy of Sciences, Baku, Azerbaijan\\
$^{c}$ Also at Novosibirsk State University, Novosibirsk, Russia\\
$^{d}$ Also at TRIUMF, Vancouver BC, Canada\\
$^{e}$ Also at Department of Physics {\&} Astronomy, University of Louisville, Louisville, KY, United States of America\\
% ...
$^{*}$ Deceased
\end{flushleft}


\end{document}